\documentclass[twocolumn]{aa}
\usepackage{epsfig}
\def\simlt{\ \raise -2.truept\hbox{\rlap{\hbox{$\sim$}}\raise5.truept   
\hbox{$<$}\ }}                                                          %
\def\simgt{\ \raise -2.truept\hbox{\rlap{\hbox{$\sim$}}\raise5.truept   %
\hbox{$>$}\ }}                                                          %
\def\newline{\hfil\break}
\def\be{\begin{equation}}
\def\ee{\end{equation}}

\begin{document}
   \title{Cooling Flows or Warming Rays?}

   \author{S. Colafrancesco
          \inst{1}
      \and
          Arnon Dar \inst{2}$^,$\inst{3}
      \and
      A. De R\'ujula \inst{3}
          }

   \offprints{S. Colafrancesco\\cola@coma.mporzio.astro.it}

   \institute{INAF - Osservatorio Astronomico di Roma,
              Via Frascati 33, I-00040 Monteporzio (Roma), Italy
         \and
             Physics Department and Space Research Institute, Technion,
             Haifa 32000, Israel
     \and
         Theory Division, CERN, CH-1211 Geneva 23, Switzerland
             }

   \date{}


\abstract{
The radiative cooling time of the X-ray-emitting plasma near the center
in many clusters of galaxies is shorter than the age of the cluster, but
neither the expected large drop in central temperature --nor the expected
mass flow towards the pressure-depleted cluster centers-- are observed.  
We propose a solution to these ``cooling flow'' problems wherein energy is
supplied to the plasma by hadronic cosmic rays (CRs). The solution
requires an energy deposition more intense and more distributed than in
conventional CR models, but this alternative model is also successful in
describing the origin of CRs, as well as the properties of gamma ray
bursts and of the diffuse gamma background radiation. The X-ray energy
emitted by clusters is supplied, in a quasi-steady state, by the hadronic
CRs, which act as ``warming rays''.  The temperature distribution in the
intracluster space is successfully predicted from the measured
plasma-density distribution. Four other puzzling features of clusters can
also be explained in simple terms: the discrepancy between their
``virial'' and ``lensing'' masses, their large magnetic fields, the
correlation between their optical and X-ray luminosities, and the
non-thermal tail of their X-ray spectrum.

 }

\maketitle

{
\section{Introduction}

The intergalactic plasma in the central regions of many clusters is cooler than
in their outskirts, and it radiates X-rays at such a rate that the plasma
cooling time is much shorter than the age of the cluster.  It has been argued
(Cowie \& Binney 1997; Fabian \& Nulsen 1977; for reviews see, e.g. 
Sarazin 1988;
Fabian 1994, 2002; Binney 2001) that in such clusters there should be a flow of
plasma from the outer hot regions to the cooler center ---a ``cooling
flow'' (CF)--- in order to maintain
hydrostatic equilibrium. For decades, the flows have been increasingly at
variance with observation. Two of the old puzzles are that the inflowing
material disappears with no observed trace and that the central-temperature
depression is not as deep as expected on the basis of the plasma's cooling rate,
with the observed central temperature $T_{inner}$ settling at a fraction (or the
order of 1/2) of the outer temperature $T_{outer}$ (see, e.g. McNamara 
1997).
More recently, high spatial resolution imaging with Chandra (e.g. McNamara et
al. 2000; Fabian 2000; Blanton et al. 2001;  Allen, Schmidt \& Fabian
2001; Lewis, Stocke \& Buote 2002; Blanton et al. 2002) and high spectral
resolution measurements with the Reflecting Grating Spectrometer of XMM-Newton
(Peterson et al. 2001, Tamura et al. 2001, Kaastra et al. 2001; Kahn et al.
2002; Peterson et al. 2002) and with XMM/EPIC (Bohringer et al.~2001a,b,c;
Molendi and Pizzolato 2001; Matsushita et al.~2002)
have sharpened the problem: lines corresponding to
gas below $T_{inner}$ are not seen, as if the gas cooled to this temperature and
then vanished.  We refer to these as the ``CF problems''.

It has been recognized that some form of heating, e.g. by a central radio
source, may alleviate the CF problems (Bohringer et al. 2002;  Churazov
et al. 2002). One difficulty with solutions along these lines is that the
heating has to be distributed, like the plasma itself is. Another
difficulty is that the bremsstrahlung cooling rate is proportional to
$n_e^2\,T^{1/2}$, with $n_e$ the plasma's electron number density: a suitable
heating and pressure-building mechanism must somehow adapt itself to this
behaviour, particularly as a function of $n_e$, which varies by orders of
magnitude along the clusters' radii\footnote{Hydrogen in the plasma is fully 
ionized, and we are using in the Introduction $n_p\approx n_e$.}.

Heating by cosmic rays (CRs) has also been discussed (Rephaeli 1987; Rephaeli
\& Silk 1995), but it does not seem to be a good solution. The traditional
theory of CRs posits that the bulk of them is accelerated by expanding supernova
(SN) shells. The total energy in CRs made this way falls short of the required
heating energy by more than one order of magnitude. Moreover, the acceleration
takes place close to the central parts of galaxies, and the CRs would not
deposit energy throughout a cluster as uniformly as it is required.

A somewhat different theory of the origin of CRs has been proposed, according to
which SNe indeed accelerate CRs, but in an unconventional way. Like
matter-accreting quasars and
microquasars, most SN explosions would be accompanied by the emission of bipolar
jets of ultrarelativistic ``cannonballs'' (CBs), with initial Lorentz factors,
$\gamma_0$, of ${\cal{O}}(10^3)$.  The asymmetry in these jets would be
responsible for the observed large ``kick'' velocity of neutron stars (Dar \&
Plaga 1999; Dar \& De R\'ujula 2000a).
When these CBs are moving close to the line of sight of an
observer, they are seen as gamma ray
bursts (GRBs) and their afterglows,  providing an extremely simple, 
detailed and successful description of
these phenomena (Dar \& De R\'ujula 2000b; Dado, Dar \& De R\'ujula 
2002, 2003a, 2003b; De
R\'ujula 2002; Dar 2003 and references therein).  The CBs travel for
kiloparsecs in the interstellar medium, decelerating by knocking out its
constituents, which are thereby accelerated to become CRs. This CR theory
provides a qualitative understanding of the CR spectrum (Dar \& Plaga 1999;
Plaga 2002) and a simple and very satisfactory understanding of the diffuse,
high-galactic-altitude Gamma Background Radiation (GBR), which originates in the
halo of our Galaxy, and is therefore {\bf not} cosmological (Dar \& De R\'ujula
2001a).  The CRs in this theory are not confined to the disk of the Galaxy, but
permeate a much larger halo.  The theory also implies a total CR production rate
that is more than an order of magnitude larger than the conventional figure, and
yet, {\bf not} in disagreement with observations (Dar and De R\'ujula 2001b).

We show here that this unconventional theory of CR production resolves in
a very simple way the CF problems and results in a predicted temperature
distribution in clusters that is in very good agreement with the
observations, while implying the absence of significant flows. What are
needed, we posit, are not cooling flows, but {\it ``warming rays''} (WRs):
cosmic rays\footnote{We do not use the expression ``warming cosmic rays'',
since cosmic rays are not ``cosmic'' in the current sense of the word.
When referring to CRs in CF clusters, we use the expressions
``CRs'' and ``WRs'' as synonyms.}
that supply the X-ray energy emitted by clusters from their central regions
and play a role in determining the temperature and pressure distribution in the
whole intracluster space.

As a bonus, our theory offers plausible ways out from four other cluster
conundra: the  X-ray/optical luminosity correlation,  the lensing/X-ray mass
discrepancy, the (quite)
large intracluster magnetic fields,
and the origin of the non-thermal tail in the  hard X-ray
spectrum:
\begin{itemize}
\item{}
The luminosities of clusters in optical and X-ray bands are observed to be
correlated: they are proportional (e.g. Miller \& Nichol 2001). Once
again, this requires an apparently surprising concatenation between two
very different emission mechanisms. This correlation, and the value of its
proportionality factor, are the ones implied by our theory of WRs.
\item{}
The mass in cluster cores estimated from X-ray measurements is
systematically a few times smaller than that estimated from observations
of gravitational lensing (e.g. Hattori et al.~1999, and
references therein). The X-ray estimate relies on an approximate
hydrostatic equilibrium of the plasma in the clusters.
If magnetic fields, WRs and the plasma they heat up
are in energy equipartition, the pressure gradient of the first two
 ---not taken into account in the assumed equilibrium--- is of the
right order of magnitude to solve the mass-discrepancy problem.
\item{}
The very large CR production rate that we presuppose corresponds
 ---again for an
assumed energy equipartition--- to the surprisingly large
fields that are observed in clusters (e.g. Carilli \& Taylor 2002; Eilek
\& Owen 2002, and references therein).
\item{}
The high energy tail of the spectrum of the X-ray emission observed in some
nearby clusters is non-thermal: $dn_\gamma/dE\sim E^{-2}$ (Fusco-Femianno et
al. 1999, 2000; Rephaeli, Gruber \& Blanco 1999). This is the expectation for
the emission from the knocked on electrons produced in the CR showers that, in
our theory, ``warm up'' the inner-cluster plasma.
\end{itemize}

We summarize in the text the features of the CB model most relevant
to the problem of CF clusters, relegating to an Appendix the outline of their
derivation, as well as a short overview of the model itself.

\section{The distribution and luminosity of CRs}

We have argued elsewhere that the ``natal kicks'' of neutron stars, cosmic 
rays,
long-duration gamma-ray bursts and the diffuse gamma background radiation have 
a common origin: highly relativistic jets of cannonballs emitted in 
supernova explosions (for CRs and neutron stars, see Dar \& Plaga 1999;  
for the GRBs and
their associated SNe see, e.g., Dado et al. 2002; De R\'ujula 2002; Dar 2003
and references therein;  for the GBR, see Dar \& De R\'ujula 2001a). In our
unconventional view of every one of these subjects, high-energy 
astrophysical phenomena would have a remarkable unity.

Of the quoted items, the most relevant to the current work are
those leading to the estimate of the relation between the optical 
luminosity and the CR luminosity of a galaxy, and to the distribution of 
CRs in space. We summarize them here.

In Dar \& Plaga 1999, Dar \& De R\'ujula 2001a,b,
and in the Appendix  we argue that
the CR luminosity of the Galaxy should be:
\begin{equation}
L_{_{CR}}
\sim 5 \times 10^{42}~\rm{erg~s^{-1}},
\label{crluminosity}
\end{equation}
which is more than one order of magnitude bigger than conventionally believed.
We also contend that the extension of the CR distribution in the Galaxy may be
much larger than generally accepted, so that the CRs permeate a  ``cosmic-ray
halo'' extending for tens of kiloparsecs. Such a CR distribution in the galaxies
of a dense cluster would permeate the cluster fairly uniformly, and result in a
CR population proportional to the number density of galaxies in the
cluster\footnote{ Active Galactic Nuclei (AGNs) emit powerful jets of CBs
that should also be efficient CR accelerators, but not all clusters 
contain observed AGNs, nor
is the CR population that AGNs generate uniformly distributed.}. For the 
O(10)  $\mu$G magnetic fields of an intracluster medium (e.g. Carilli \& 
Taylor 2002), the
bulk of the CRs does not escape the cluster over times comparable to its age
(e.g. Colafrancesco \& Blasi 1998).

Our arguments are based on observations of
{\it long-duration} GRBs and {\it core-collapse} SNe (of type II, Ib and 
Ic). In using them to discuss the CR luminosity of clusters we ought
to ponder the relative frequencies of SNe
of various types in spirals or ellipticals, and on the abundances
of the various galaxy types in clusters. Though such considerations
would not affect our results by large factors,
we make the simplifying assumption that all
SN types produce jets and CRs of
similar energy:
Type Ia supernova explosions occur as the
mass of a white dwarf in a binary increases beyond the
Chandrasekhar limit because of the mass accretion from a
non-degenerate companion or a merger with another white dwarf.
These processes result in an unstable accretion configuration
akin to that observed in micro-quasars, and should also
result in the emission of jets. If, as we presume, Type Ia SNe
are associated with {\it short-duration} GRBs, the energy of
their jets should not be much smaller than for core-collapse SNe.

The contention that long-duration GRBs are produced by
core-collapse SNe (Dar 1999, Dar \& Plaga 1999, Dar and De
R\'ujula 2000, Dado et al. 2002) has recently received striking
observational support: the predictions of Dado, Dar\& De R\'ujula 2003c of
a SN associated with GRB 030329, its properties and even
{\it the date when it would be convincingly discovered} have
been found to be correct (Stanek et al.~2003).

\section{Cosmic rays in the intracluster (IC) medium}

The IC space of dense clusters is filled with a highly ionized
plasma at a
temperature $T$ of a few keV, electron
number density $n_e\!=$ $10^{-2}$ to
$10^{-3}$ cm$^{-3}$ near the center,
and slightly sub-solar metal abundance.
This plasma is
responsible for most of the cluster's thermal X-ray emission, via
bremsstrahlung in the 0.1 to 10 keV range (see, e.g. Sarazin 1988).

In the CB model the
main acceleration mechanism of electrons and nuclei by CBs is simply the
``collisionless'' scattering of these ambient particles by the magnetic field of
the moving CBs, which makes them recoil with energies proportional to their
masses. Thus, the ratio of electron to nuclear CR energies is $\sim m_e/M$,
with $M$ the mass of the corresponding nucleus:
the bulk of the CR energy is carried by CR nuclei, mainly protons.

The dominant energy-loss mechanisms for hadronic CRs in the IC medium are
Coulomb collisions with the ambient electrons, which lead to
electromagnetic showers initiated by the {\it knocked-on} electrons, and
hadronic collisions with the ambient nuclei, which also lead to
high-energy electrons and positrons via pion production, followed by the
decay chains $\pi^{+}\to \mu^{+}\,\nu$, $\pi^{-}\to \mu^{-}\,\bar\nu$;
$\mu^{\pm}\to e^{\pm}\,\nu\,\bar\nu$. A relatively small fraction of the
available energy escapes, carried away by neutrinos and by $\gamma$ rays
from $\pi^0$ production and decay. The rest of the CR energy is deposited
in the plasma and thermalizes, via electromagnetic interactions of the
electrons and positrons, and elastic collisions of the low energy hadronic
CRs with the IC electrons and nuclei.

A highly ionized  plasma is much more efficient than
non-ionized gas in slowing down the CR nuclei.
Energy losses via Coulomb
collisions with neutral atoms become significant only for energy
transfers larger than their ionization potential $I$. The corresponding
small impact parameters imply relatively
small cross sections. For the IC plasma the role of $I$ is played
by the much smaller quantity $I_p\!=\!\hbar\, \omega_p$, with
$\omega_p\!=\! [4\pi\,  n_e\,e^2/m_e]^{1/2}$ the plasma frequency.
Let $Z^2$ 
be the (suitably averaged) squared charge
of the plasma's nuclei. For moderately relativistic CR nuclei
with a charge $z$ moving
with a velocity $v\!=\!\beta\, c$ and Lorentz
factor $\Gamma$, the energy-loss rate is:
\begin{equation}
-{dE\over dt}\approx K \, z^2\, Z^2\,{1\over \beta}\,
   \left [ln {2\, m_e\, c^2\, \beta^2 \Gamma^2\over I_p}
 -\beta^2\right],
\label{dedx}
\end{equation}
with  $K\!=\!4\, \pi\, n_e\, r_e^2\, m_e\, c^3$ and $r_e^2\!=\!e^2/m_e\,
c^2\!\simeq\!2.82$ fm. The rate of Eq.~(\ref{dedx}), for
typical plasma densities near the center of clusters
($n_e \!\sim\!10^{-2}$ cm$^{-3}$), is $\sim\! 4$
times larger than for neutral hydrogen. For protons ($z\!=\!1$) with
a typical CR energy $(\Gamma\!=\!2)$ and for a plasma consisting of
$\sim\! 0.93\%$
H, $\sim\! 0.07\%$ He (by number), and traces of heavier
elements, $Z^2\!\sim\! 1.14$
and $dE/dt\approx 10^{-17}$ GeV/s, for the quoted electron density.

Hadronic collisions between CRs and plasma nuclei constitute an
energy-loss mechanism approximately as efficient as that of Eq.~(\ref{dedx}).
Consider the collisions of CR protons with ambient protons, the discussion
of proton--nucleon and nucleon--nucleon collisions being far too cumbersome 
to be justified by the small admixtures of nuclei heavier than protons
in the CR and IC constituencies.
For $\Gamma>2$ of the incident proton the $pp$ cross section is
$\sigma_{pp}\!\sim\! 40$ mbarn, and dominantly inelastic. In practically all of these
collisions the incident proton survives unscathed, as the ``leading'' (most energetic)
final-state particle, carrying on average some 70\% of the incident proton's
energy. Thus, a proton's energy-loss rate by hadronic collisions in a hydrogenic
plasma is:
\begin{equation}
-{dE\over dt}\sim 0.7\,n_e\,\sigma_{pp}\,c\;E;
\label{dedx2}
\end{equation}
that is $dE/dt\!\sim\!1.6\times 10^{-17}$ GeV/s, for our reference $n_e$
and $\Gamma\!=\!2$. This is very close to the Coulomb energy loss
computed at the end
of the previous paragraph for the same reference parameters.

The characteristic time for a CR proton to lose its energy is $\tau=E/(dE/dt)$,
with $dE/dt$ the sum of Eqs.~(\ref{dedx}) and (\ref{dedx2}).
Once again for the chosen reference values,
 $ \tau \!\sim\! 4 \times 10^{16}$ s. This
is shorter than the age of the cluster which, for the relatively nearby 
clusters of interest here, is approximately the age of the Universe,
$t_U\!\sim\!3\times 10^{17}$ s.

To summarize, to a good approximation, CRs deposit their energy within
a CF cluster in a time shorter than the cluster's age.
Thus, in the absence
of other significant energy sources, the IC plasma cools until its
energy-loss rate by bremsstrahlung
is balanced by the CR energy-deposition rate.  Since the source of
X-ray energy at the cluster centers are the ``warming''
CRs, the cluster luminosities in CRs and
X-rays are approximately equal: $L_{_{CR}}\!\simeq\!L_{_{X}}$.

\section{The X-ray luminosity of a cluster}

``Cooling Flow'' clusters contain large amounts of plasma and,
as we have argued, we may
estimate their X-ray luminosity as the CR luminosity of the
galaxies they contain. The latter, in turn, can be estimated from
their optical luminosity, $L_\star$, as follows.

The Shapley--Ames ``fiducial'' sample of 342 galaxies within the
Virgo circle has a mean-galaxy B-band luminosity
$L_{_{\langle G\rangle}}\!=\!6.7\times 10^9\, L_\odot(B)$
and, for a galaxy or cluster of B-band luminosity $L$,
a mean rate of $2.07\,L/L_{_{\langle G\rangle}}$
SNe per century
(Van den Bergh and Tamman 1991).  We have
argued that the total CR energy produced by a SN is roughly twice
the jet energy estimated from the typical neutron-star velocities;
see the Appendix, Eq.~(\ref{Ejet}).  Assembling
these facts and converting from a
B-band to an optical solar reference, we conclude that the X-ray, CR
and optical luminosities of clusters ought to satisfy:
\begin{equation}
L_{_{X}}\approx L_{_{CR}}\approx 0.02\, L_\star.
\label{lumcr}
\end{equation}
The predicted relation between $L_{_{X}}$ and $L_\star$
is consistent with the observations for CF clusters, as we show
in Fig.~(\ref{lum}), adapted from Miller et al. 1999.
The panel (c) in this figure disagrees with the prediction,
but it refers to groups and not clusters of galaxies. Since
 WRs are not confined in the magnetic field of a group,
their X-ray and optical luminosities should be linearly related, albeit with a
smaller coefficient than that of Eq.~(\ref{lumcr}), as observed. Encouraged by
these results on the optical/X-ray luminosity relation, we proceed to
study its radial distribution in CF clusters.


\section{The temperature profile in cooling flow clusters}

The ``virial'' expectation for the temperature of the plasma
in a cluster, for a spherical distribution with mass
$M(r)$ within a radius $r$, is:
\begin{eqnarray}
 k\,T(r)&\simeq& {G\, m_p\, M(r)\over 3\, r}\nonumber\\
&\simeq&
          2\;{\rm keV} \left[{M(r)\over 10^{13}\, M_\odot}\right]\,
          \left [{r\over 100\, {\rm kpc}}\right]^{-1}\, .
\label{Tvirial}
\end{eqnarray}
The observed ``outer'' temperature of the plasma in a cluster, $T_{outer}$, is
---at large radii where the cooling time should be longer than the cluster's
age--- roughly compatible with the virial expectation. In the inner regions the
temperature is somewhat smaller ($\sim 1/2$ of $T_{outer}$) than that of
Eq.~(\ref{Tvirial}), and significantly larger than the cooling rate would imply
(see, e.g., Peterson et al. 2001).

What is the effect of the ``warming rays'' on a cluster's temperature
distribution? To answer this question we make the following simplifying
assumptions:
\begin{itemize}
\item{} Spherical symmetry.
\item{} The density of the plasma is proportional to that of galaxies
and of the SNe they contain (clearly without
sufficient ``resolution'' to individuate single galaxies).
This implies that the CR distribution ---which in our model
traces with bad resolution
the SN distribution--- also traces the plasma distribution: {\it the radial
distribution of the CR
intensity is proportional to that of the plasma density.}
\item{} The temperature of the plasma in its ``initial state'' (as the
cluster is born) can be approximated by a radius-independent
constant.
\item{} Cooling by X-ray emission and heating and pressure-building
by WRs are the dominant evolutionary agents. Flows and cluster ageing
are relatively unimportant.
\end{itemize}

Given our assumptions, the
rate per unit volume at which the plasma loses energy
by the emission of X-rays is:
\begin{eqnarray}
&&{d\epsilon\over dt}\Biggr|_{_{X}}= a\, [n_e(r)]^2\, \sqrt{T(r,t)};\nonumber\\
&&a= \sqrt{2^{11}\pi^3\over 3^3}\;
{e^6\sqrt{m_e }\over h\,m_e^2\, c^3}\, g\,
{\bar z}\nonumber\\ &&\sim
4.8\times 10^{-24}\,{\bar z}\,{1\over \sqrt{\rm keV}}
{\rm erg\,cm^3 \over  s}\; ,
\label{Xloss}
\end{eqnarray}
where $\bar z$ is an average charge of the IC matter and in the
numerics we have approximated the Gaunt factor $g$ by unity.
The rate per unit volume at which the WRs deposit energy
in the plasma is also quadratic in $n_e$:
\begin{equation}
{d\epsilon\over dt}\Biggr|_{_{WR}}=
b\, [n_e(r)]^2,
\label{WRgain}
\end{equation}
since we have first argued and then assumed that the WR production 
rate per unit volume is proportional
to $n_e$. The evolution of the temperature anywhere in the IC plasma 
satisfies the energy-conservation relation:
\begin{equation}
3\,k\, n_e(r)\, {dT(r,t)\over dt}=
{d\epsilon\over dt}\Biggr|_{_{WR}}
 - {d\epsilon\over dt}\Biggr|_{_{X}}\, ,
\label{Tevol}
\end{equation}
with the initial condition $T(r,0)=T_i$.
Rewrite Eq.~(\ref{Tevol}) as:
\begin{equation}
-{dT(r,t)\over  \sqrt{T} - b/a}= {a\, n_e(r)\over 3\, k}\, dt
\label{dT}
\end{equation}
and integrate, to obtain:
\begin{eqnarray}
ln \left[ {\sqrt{T_i/T_f}-1\over \sqrt{T(r,t)/T_f}-1}\right
]-\sqrt{T(r,t) \over T_f}+\sqrt{T_i \over T_f}=
 \nonumber \\
{b\, n^2_e(r)\over 6\,n_e(r) k T_f}\, (t-t_i)
 \label{Tr2}
\end{eqnarray}
where $T_f\equiv (b/a)^2$ is the steady temperature first reached
at the center of the cluster, where the density is high, and
at asymptotically late time in the periphery, where the density is vanishingly
small. We have written the r.h.s. of Eq.~(\ref{Tr2}) in a way that
shows that the evolution of $T(r,t)$ is determined by
the ratio between the WR deposition rate $b\, n_e^2$
and the thermal energy per unit volume at the cluster's central temperature.

The temperature evolution described by Eq.~(\ref{Tr2}) is shown in
Fig.~(\ref{fig.cfevolution}), for $t=0,t_{_{U}}/3,t_{_{U}}$ and $3\,t_{_{U}}$,
with $t_{_{U}}$ the current age of the Universe\footnote{We use a
cosmology with $\Omega_\Lambda=0.7$, $\Omega_M=0.3$
and $H_0=75\, km\, s^{-1}\, Mpc^{-1}\, $ but in relating the radial 
coordinate
of a given cluster to the observed angular coordinate, we use in
each case the value of $H_0$ assumed by the observers.}. In the particular 
example in Fig.~(\ref{fig.cfevolution}) the central density is high enough
for the equilibrium between X-ray cooling and WR heating to have
been reached at the center of the cluster in a time of the order of
its current age. In such a case, $T_f\simeq T_{inner}$, the presently
observed inner-cluster temperature. We shall see that some clusters
have reached this stage, while others have not.
For all the clusters we study, the cooling time in the outer
parts of the cluster is much longer than $t_{_{U}}$, so that
$T_i\simeq T_{outer}$.

\section{Temperature distributions of specific clusters}

We fit Eq.~(\ref{Tr2}) to the temperature distribution of a given cluster as
follows: we fix from the observations an ansatz value of $T_i\simeq T_{outer}$, 
we input the observed $n_e(r)$ (either point by point, or via a ``$\beta$-model''
or a variation thereof) and we vary a single parameter $b$ ---or, equivalently,
$T_f$--- to obtain a best fit $T(r,t)$ at the current age of the clusters
($t=t(z)$ is fixed by  the cosmological model we have adopted).

We show in Figs.~(\ref{fig.1795}) to (\ref{fig.2029}) the
measured density profiles and the predicted and observed
temperature profiles of the clusters A1795, Sersic 159-03,
Hydra A, and A2029. The temperature and density data are
taken, respectively, from Tamura et al. (2001), Kaastra et al.
(2001), David et al. (2001) and Lewis et al. (2002).
These figures demonstrate that the correlation between
the input density profile and the output temperature
profile predicted by Eq.~(\ref{Tr2}) is in excellent agreement
with the observations: in all cases the shape of
$T(r)$ is snugly reproduced.

The examples of A2052 and EMSS1358+6425 are shown in
Figs.~(\ref{fig.2052})
and (\ref{fig.emss}). The data are from Blanton et al. (2002) and
Arabadjis et al. (2001), respectively. In the first of these
clusters, the observed $n_e(r)$ is not very smooth and we have made
two predictions for $T(r)$. One prediction uses, as in all other cases, 
the best fit $n_e(r)$ and results in a smooth prediction for 
$T(r)$.
The other prediction uses the measured point-by-point density, and
results in a point-by-point prediction for $T(r)$, shown in
Fig.(\ref{fig.2052}) as a series of triangles. The latter prediction
follows the shape of the observed $T(r)$ somewhat better than the
smooth curve does.
EMSS1358+6425 is special, in that its measured
$T(r)$ is compatible with a constant, albeit with very large errors.

The values of the central asymptotic temperature, $T_f$, resulting from our fits
are shown in Table I, and compared with $T_{inner}$, the innermost
observed temperature. Also given in the table are the values of $T_i$, the 
initial
constant input temperature, and of $T_{outer}$, the outermost observed
temperature\footnote{For two clusters, Sersic 159-03 and A2052, the quoted
value of $T_{outer}$ is that of the penultimate radial point, since
the lower value of the last point may indicate the ``end'' of the cluster.}.
The meaning of all these temperatures is illustrated in 
the lower panel of Fig.~(\ref{fig.cfevolution}). In
Sersic~159-03 and in Hydra, $T_f=T_{inner}$ within errors, so that the 
centers of the cl;usters have reached a nearly steady final temperature.
For A1795 and A2052 there is an indication that $T_f\!<\!T_{inner}$,
so that the cluster's central temperature should still
cool in the future (not a prediction the success of which we are likely to
witness). In these four clusters, $T_{outer}\!<\! T_i\, ,$ indicating
that their outer regions have also cooled with time.
The fits to the temperature profile in Figs.~(\ref{fig.1795}) to (\ref{fig.2029}) 
are very satisfactory.
In the case of A2052 the observed density is so erratic that
the formal errors of our simple fit cannot be trusted; 
the large $\chi^2$ reflects this, as is
clear in Fig.~(\ref{fig.2052}). For EMSS1358+6425, shown in
Fig.~(\ref{fig.emss}), the errors in the observed temperatures are
so large that the fit is not very illuminating.

\section{The mass discrepancy in clusters}

The gravitational lensing of distant galaxies 
by an intervening cluster can be used to infer the cluster's
mass distribution
(e.g.  Grossman \& Narayan 1989).  Such mass estimates are quite consistent with
the ones derived from the observed velocities of the galaxies in the clusters.
On the other hand, the mass deduced from lensing or from the motion of the
galaxies is $\sim\! 3$ times larger than the mass deduced from the clusters'
X-ray emission (e.g. Hattori et al. 1999). Fabian and Allen (2003) obtain an
agreement between the two mass estimates by using the density profile 
obtained by
numerical simulations (Navarro, Frenk \& White 1997), but this profile deviates
significantly from the observed plasma distribution near the center (where it 
diverges as $1/r$) and in the periphery (where its $1/r^3$ asymptotic behaviour 
is significantly steeper than the observed one).

Assume the WRs to be in approximate energy equipartition
with the magnetic field that confines them. Given our other assumptions
on WRs, their pressure is proportional to the thermal pressure of
the plasma. Let $\Phi(r)$ be the gravitational potential of a
spherically symmetric cluster
and $\mu$ the mean atomic weight of its plasma.
The total mass-density distribution of the
cluster, $\rho_{tot}$, can be inferred from the density distribution of the
X-ray-emitting gas, $\rho_{gas}$, via Poisson's
equation\footnote{We neglect the (logarithmic) temperature
gradient relative to the corresponding density gradient, a good
approximation in the cases at hand.}:
\begin{eqnarray}
\rho_{tot}&=&{1\over 4\, \pi\, G} \nabla^2 \Phi \nonumber \\
& = & {(1+\alpha)\, k\, T
            \over 4\, \pi\, G\, \mu\, m_p}\, \nabla^2\, ln\,
            \rho_{gas}^{-1}\, ,
\label{Poisson}
\end{eqnarray}
where $\alpha$ is the proportionality constant relating the sum of the magnetic
and CR pressures to the thermal pressure. In approximate equipartition
($\alpha\simeq 2$), the total mass deduced from the temperature and density
distribution of the X-ray-emitting plasma is approximately three times larger
than that inferred by neglecting the magnetic and CR pressures. This 
brings 
the
different mass estimates to better agreement, as can be seen from
Fig.~(\ref{mass}), adapted from Hattori et al. (1999). The substantial
contribution of the magnetic and WR pressures helps in explaining why our last
simplifying assumption in Section 5 ---that flows are relatively unimportant in
determining the temperature profiles--- may be a fair approximation.

If the magnetic field is in rough energy equipartition with the
WRs, the corresponding pressures are related by $B^2/(8\pi)=2\,n_e\,k\,T$.
The magnetic field would thus be:
\begin{equation}
B(r)\sim (20\,  \mu {\rm Gauss})\;\left[{T(r)\over 3\,{\rm keV}}\right]^{1/2}\;
\left[{n_e(r)\over 10^{-2}\,{\rm cm}^{-3}}\right]^{1/2},
\label{B}
\end{equation}
which is in the observed domain. (e.g. Carilli and Taylor 2002; Eilek \&
Owen 2002; and references therein).

\section{The X-ray tail}

CR nuclei of charge $z$, velocity $\beta\,c$
 and Lorentz factor $\Gamma$ deposit their
energy in the IC plasma via Coulomb scattering of the plasma
electrons. These knocked-on
electrons recoil with kinetic energies up to $K_{max}\approx
2\, m_e\, c^2\, \Gamma^2$.  In a plasma of electron density $n_e$
the number of electrons per unit time and volume scattered to a
kinetic energy $K$ in the range  $\hbar\,\omega_p \ll K \leq K_{max}$
is:
\begin{eqnarray}
{d^2 N_e\over dK\, dt}&\approx&
2\, \pi \,r_e^2\, n_e\, m_e\, c^3\, z^2 \;{F(K)\over \beta\, K^2}\, ,
\nonumber\\
 F(K)&\approx& 1-\beta^2 {K\over K_{max}}\, .
\label{dndK}
\end{eqnarray}
We are interested in electrons of energy 
much smaller than the $K_{max}$ values corresponding to typical CRs, so that the
integration over the CR spectrum, which peaks at
$\Gamma\!-\!1\!=\!{\cal{O}}(1)$, preserves the $K^{-2}$ shape of
Eq.~(\ref{dndK}) up to $K\!=\!{\cal{O}}(m_e\, c^2)$.

The scattered electrons cool and thermalize in the plasma mainly
by bremsstrahlung emission at high energy and by Coulomb collisions
at low energy. A detailed calculation of the resulting quasi-steady
electron energy distribution is beyond the scope of this paper,
but this distribution must be thermal at low energies with a
power-law tail whose power-law index is $n\!\approx\! 2$.  Hence, the
thin bremsstrahlung emission, which dominates the radiative cooling
of these electrons, has a thermal (thin) bremsstrahlung shape
 at low energies, with a power-law
tail with the same index as the energetic electrons.
 This behaviour (prior to photoabsorption corrections along the
line of sight) may be roughly interpolated by:
\begin{equation}
{dn_\gamma\over dE}\propto E^{-1}\,e^{-E/T}+\delta\,E^{-n}\; .
\label{dndt}
\end{equation}
Such ``thermal'' distributions with the expected power-law tail have been
observed in the X-ray emission of some clusters, e.g. by BeppoSAX (e.g.
Fusco-Femiano et al. 1999; 2000) and by RXTE (Rephaeli, Gruber \& Blanco, 1999).
Non-thermal power-law tails in the ``thermal'' bremsstrahlung emission 
from
plasmas that contain high-energy CRs have also been observed in SN 
remnants
(e.g. Mineo et al. 2001; Dyer et al. 2001), elliptical galaxies (e.g., 
Guainnazzi \& 
Molendi,
1999), and GRBs (e.g. Dar \& De R\'ujula 2000b; De R\'ujula 2002; Dar 2003
and references therein).

The example of the spectrum of the SN remnant SNR MSH 15-12 (Mineo et al. 2001)
is given in Fig.~(\ref{spectrum}), showing that the data can be described by
Eq.~(\ref{dndt}) with $n=2$, as expected, and (in keV units) $T=5$ and $\delta=1.5$.
The figure also shows the spectrum of the cluster A2256 (Fusco-Femiano et
al. 2000), for which we used $n=2$, $T=5$  and $\delta=3$.
A relatively large $\delta$ in a given cluster
is indicative of a
large WR supply, so that its central temperature trough should be
relatively unpronounced. There are currently no data to test this
expected correlation.

We conclude that the observed non-thermal power-law tail of the X-ray emission
from CF clusters is but one example of the behaviour of plasmas subject to a
high-energy CR flux, and that its presence adds consistency to our contention
that the cluster's plasma is subject to an intense flux of ``warming 
rays''.

\section{The radio emission}

The diffuse  radio emission from the halo or IC medium of clusters was first
detected by Large et al. (1959) in Coma, and first studied by Wilson (1970). It
is due to synchrotron radiation in the magnetic field of the IC medium, emitted
by electrons with  Lorentz factors $\Gamma_e\sim 10^{4\pm 1}$. A detailed
discussion of this radio emission is beyond the scope of this paper, but some
comments are appropriate.

The radio-emitting electrons lose energy by inverse Compton
scattering, with a cross-section
$\sigma_T\!=\!0.665\, \times 10^{-24}$ cm$^{-2}$,
 on the real photons of the cosmic background radiation
and on the  virtual ones of the cluster's magnetic field, whose energy
densities are $\epsilon_{_{CMB}}\!\simeq\! 0.26$ eV cm$^{-3}$ and
$\epsilon_{_{B}}\!=\!B^2/(8\pi)$. The  electron's cooling time:
\begin{eqnarray}
\tau_e&=&{3\, m_e\, c \,  \over 4\, \sigma_T\,
(\epsilon_{_{CMB}}+\epsilon_{_{B}})\,\Gamma_e}
\nonumber\\
&\sim& (7.4\times 10^{15}\, {\rm s})\, {\epsilon_{_{CMB}}\over
 \epsilon_{_{CMB}}+\epsilon_{_{B}}}
\, {10^4\over\Gamma_e}\; ,
\label{ecool}
\end{eqnarray}
is very much shorter than the age of the cluster. Consequently, these 
electrons
must be supplied continuously and fairly uniformly over the cluster. In
particular, they cannot be supplied by electron-accelerating SN shells, since
the electrons' cooling time is also much shorter than their diffusion time
out of the dense regions of galaxies.

In our WR model, the high-energy electrons are supplied
---steadily and in a distributed fashion---
by the highly relativistic and far-reaching CBs jetted by SNe (and AGNs). The
energy of these electrons is locally re-emitted by synchrotron radiation and
dominates at radio frequencies.

\section{The $\gamma$-ray emission}

In a cluster, as we have shown,
the CR protons lose about as much energy to knocked-on
electrons as they do in hadronic collisions with the IC nuclei. Very roughly,
the CR proton spectrum carries comparable kinetic energy above and below
1 GeV. The higher-energy protons, when colliding hadronically, transfer
about 1/3 of their energy to $\pi^0$'s, which decay into $\gamma$ rays.
This implies that the $\gamma$-ray luminosity of a CF cluster ought
to be (very) roughly 1/6 of its X-ray luminosity.

The $\gamma$-ray spectrum
is hard to ascertain with confidence, one reason for this being that the 
spectral shape
of the parent CRs in a cluster is not known. The spectral shape of the
CRs on Earth is affected at low energies by the planet's magnetic field
and, at all but the highest energies, by the energy-dependence of the
confining time of CRs in the Galaxy and its halo. Predicting the spectral
shape of a cluster's $\gamma$-ray spectrum, therefore, requires extra
assumptions beyond the ones made in this paper.

\section{Conclusions}

Cooling flow clusters were named after an inference ---the flow itself---
and not a direct observation. It is therefore not very surprising that further
observations led to contradictions.
We have seen how an extremely simple model of the origin, luminosity
and location of cosmic rays provides a good understanding of the
main ``cooling flow problem'': the observed temperature
profile of these clusters. This {\it Warming Ray} model
also explains in simple
terms the clusters' X-ray luminosity, its correlation
to the optical luminosity, the cluster's magnetic field intensity, the discrepancy
between virial and lensing masses, and the high-energy tail of the X-ray spectra.

The Warming Ray model has no specific consequences regarding other
interesting cluster properties ---such as their ``metal'' abundances--- but it
is part of a non-standard unified view of high-energy astrophysical
phenomena, which includes the ``natal kicks'' of neutron stars, quasar and
microquasar emissions, $\gamma$-ray bursts and their afterglows, and the
origin and acceleration of cosmic rays.

\section{Appendix: Astrophysical Cannonballs}

We briefly review the CB model of GRBs, as well as
our unconventional views on the CR distribution
and luminosity of galaxies (for other reviews and
for references, see e.g. De R\'ujula 2002, Plaga 2002, Dar 2003).

\subsection{The cannon}

The ``cannon'', or engine, producing the jets of CBs, is not
understood. We assume that, in a SN explosion, a small
fraction of the infalling matter may be
jetted along the axis with highly relativistic velocities
in a succession of shots, before its reservoir is exhausted. This
emission of CBs would be akin to that observed in quasars and
microquasars, in which episodes of accretion from a disk or torus into a
central compact object result in the axial ejection of relativistic
``plasmoids'' ---or CBs--- made of ordinary matter\footnote{In the CBs of
the micro-quasar SS 433, Balmer H and He lines (e.g. Eikenberry et
al. 2001) and the K$\alpha$ line of Fe (Migliari et al. 2002) have been
observed.}.

\subsection{The GRB}

When crossing the SN shell and the matter distribution
produced by the ``wind'' emitted by the progenitor star, the front
 of a CB is heated to a time-dependent
temperature $T$, of ${\cal{O}}(1)$ keV. The typical initial Lorentz
factor of a CB, $\gamma_0\!\sim\! 10^3$, follows from the observed
energies of the GRB photons which, Doppler-boosted and cosmologically
red-shifted, peak at a fraction of 1 MeV.  The quasi-thermal
radiation a CB emits, boosted and collimated by its relativistic
motion, is
a single $\gamma$-ray pulse in a GRB.  The cadence of pulses reflects
the chaotic accretion and is not predictable, but the individual-pulse
temporal and spectral properties are.

\subsection{The afterglow of a GRB}

The ejected CBs, as observed in micro-quasars, are assumed to contain a
tangled magnetic field.  As they plough through matter, they gather and
magnetically scatter its baryonic constituents, mainly protons. The
re-emitted protons exert an inward pressure on the CBs, which counters
their expansion and makes them reach an asymptotic radius $R_{_{CB}}$, in
minutes of the observer's time\footnote{This mechanism may explain the
surprisingly small size of the CBs emitted, for instance, by the Pictor-A
quasar (Wilson, Young, \& Shopbell 2001)}.  The GRB's afterglow is
dominated by synchrotron radiation by the electrons swept in by the CB in
its voyage through the interstellar medium (ISM).


\subsection{The generation of CRs}

A CB ploughing through the ISM decelerates, and
its time-dependent Lorentz factor $\gamma(t)$ is explicitly calculable.
The ISM nuclei of mass $M$ that are simply reflected by the CB's
magnetic field at time $t$ recoil with a flat energy distribution
extending to $2\,\gamma(t)^2\,M\,c^2$. Their time-integrated spectrum
is the CR spectrum, extending up  to the spectral feature known as
the ``knee''. The higher-energy spectrum would be due to the ISM nuclei
that have been Fermi-accelerated in the CB's tangled magnetic field before
they exit again into the ISM. The overall CR spectrum and the interpretation
of its very high-energy ``ankle'' are discussed in Dar \& Plaga 1999.
The fraction of the original CB's energy ending up in electrons
is $\simlt m_e/m_p$. Thus, the overall efficiency of this
mechanism in accelerating CR nuclei is $\sim 100$\%: essentially
all the initial kinetic energy of a CB ends up in CR nuclei.

\subsection{The CR luminosity of the Galaxy}

What is the total energy of the CRs produced by the jets of CBs
emitted in a single SN explosion? In the case of a core-collapse SN
leading to the formation of a NS, we may roughly
estimate this quantity as follows.
The mean sky velocity of neutron stars (Lyne and Lorimer, 1964) is
$ \langle v_{_{NS}}\rangle\simeq
450\pm 90$ km s$^{-1}$.
Assume this velocity to be due to a ``natal kick'': a
momentum imbalance between the two jets of CBs emitted axially
in opposite directions, as the NS is born.
This implies:
\begin{equation}
E_{jet}\!>\!M_{_{NS}}\, v_{_{NS}}\, c\simeq 4\times
10^{51} {M_{_{NS}}\over 1.4\, M_\odot} \;\rm erg,
\label{Ejet}
\end{equation}
with the inequality becoming an equality in the case
of a large momentum asymmetry.
The total energy per SN ending up in CRs
would be of the order of twice the r.h.s. of Eq.~(\ref{Ejet}).
1;2c1;2c
The estimated rate
of Type II, Ib and Ic supernovae in the Galaxy is $ R_{_{SN}}\sim  1/50$
per year (van den Bergh and Tammann 1991).
The CR luminosity
of a Milky-Way-equivalent galaxy would then be given by:
\begin{equation}
L_{_{CR}}\sim 2\, E_{jet}\, R_{_{SN}}
\approx 5.1 \times 10^{42}~\rm{erg~s^{-1}},
\label{crluminosity2}
\end{equation}
the numerical result being the one quoted in Eq.~(\ref{crluminosity}).
This is more than one order of magnitude larger than the classic
estimate $L_{_{CR}}\sim 1.5\times 10^{41}$ erg s$^{-1}$, based on
the observed ratios of primary CRs to the secondary ones produced
by collisions with the ISM (Drury et al.~1989).
 The alterity can be easily understood: the ISM gas density, volume
and grammage used to derive the classic result all refer to a region
close to the visible Galaxy.  The derivation is invalid (e.g.
Longair 1992) if the locally observed CRs may have spent a fraction
of their travel-time in a much less dense, though magnetized halo.

\subsection{The GBR and the locus of CRs}

In Dar \& De R\'ujula 2001a we analysed the possibility that the
diffuse Gamma Background Radiation
at high galactic latitudes could be dominated by inverse
Compton scattering of CR electrons on the cosmic microwave background
radiation, and on starlight from our own galaxy.  Assuming
the mechanisms that accelerate galactic CR hadrons and electrons to
be the same ---as they are if accelerated by CBs--- we derived
simple and successful relations between the spectral indices of
the GBR above a few MeV, and of the CR electrons and CR nuclei
above a few GeV. We reproduced the observed intensity and angular
dependence of the GBR, in directions away from the galactic disk
and center, without recourse to hypothetical extragalactic sources:
the GBR is, we argued, predominantly not ``cosmological''.

The above results require that the CR population of the Galaxy extend 
well beyond
its disk. In a model with a fitted gaussian scale height $h_e\sim 20$ kpc above
the Galactic plane, and a scale radius $\rho_e\sim 35$ kpc in directions
perpendicular to the Galactic axis, we reproduced the observed properties of the
GBR. To obtain these results, we assumed the nuclear CRs and the CR 
electrons to 
be equally distributed ---as they would, if accelerated by CBs---, with their
flux ratio fixed to the locally observed value.  A scale height of CR nuclei as
high as $h_{_{CR}}= h_e\sim$ 20 kpc rises the eyebrows of some experts and the
swords of referees, but is not excluded by data on relative CR
abundances\footnote{For the most elaborate models (Strong 
and Moskalenko 1998) a
``leaky-box'' scale height of 20 kpc is only some 1.3$\sigma$ above the 
central
value of the most precise observations (Connell et al. 1998); it is 
perfectly
compatible with the average of all previous and somewhat less precise results,
1;2c1;2c1;2ccompiled in Lukasiak et al. 1994.}.
A CR distribution of this large scale is a ``cosmic-ray halo''. Its computed
total luminosity, for the fitted values of $h_e$ and $\rho_e$, coincides with
the ``large'' value of Eq.~(\ref{crluminosity}).

A CR halo of similarly distributed electrons and ions is what is expected if CRs
are accelerated by CBs, emitted by SNe close to the galactic disk.  A CB of mass
$M_{_{CB}}$ and roughly constant cross section $S_{_{CB}}$, travelling in 
an
interstellar medium of roughly constant density $\rho_{_{ISM}}$, has its Lorentz
factor diminished from $\gamma_0$ to $\gamma$ in a distance:
\begin{equation}
x=x_\infty\left[{1\over \gamma}-{1\over \gamma_0}\right],
\label{x}
\end{equation}
where $x_\infty\!=\!M_{_{CB}}/(\rho_{_{ISM}}\,S_{_{CB}})$. By explicitly fitting
the spectra and time dependence of the afterglows of all GRBs of known 
redshift,
Dado et al. (2002, 2003a) have found that the fitted values of $x_\infty$ are
distributed over two orders of magnitude, peaking at $x_\infty\sim 1$ Mpc, while
the values of $\gamma_0$ are narrowly peaked around $10^3$. It takes a distance
$x_\infty/\gamma_0\sim 1$ kpc for a CB to half its original Lorentz factor,
while transferring half of its energy to CRs.  We cannot extend with confidence
Eq.~(\ref{x}) all the way to $\gamma\sim 1$ and trust the result: afterglow
fluences decrease with a large power of $\gamma$ and cannot be followed long
enough, and one does not expect the approximations of constant $S_{_{CB}}$ and
$\rho_{_{ISM}}$ to hold over distances much longer than a few kpc. Be it as it
may, Eq.~(\ref{x}) indicates that it takes very much longer for a CB to lose and
transfer the other half of its kinetic energy: ${\cal{O}}(1)$ Mpc  in the stated
untrustworthy approximations.

\clearpage
\begin{table}[htb]{}
\vspace{2cm}
\begin{center}
\begin{tabular}{|*{6}{c|}}
\hline
 Cluster & $z$ &  $T_f$  & $T_{inner}$ & $T_{outer}$ & $T_i$ \\
         \hline
 A1795          & 0.0625  & $3.18^{+0.14}_{-0.17}$ & $3.47^{+0.13}_{-0.13}$ &
                  $6.2^{+0.4}_{-0.4}$ & $6.6$ \\
 Sersic   & 0.0564  & $2.28^{+0.16}_{-0.15}$ & $2.38^{+0.09}_{-0.09}$ &
                  $2.69^{+0.13}_{-0.13}$ & $2.8$ \\
 Hydra          & 0.0539  & $3.18^{+0.23}_{-0.22}$ & $3.11^{+0.14}_{-0.14}$ &
                  $3.89^{+0.33}_{-0.33}$ & $4.0$ \\
 A2029          & 0.0767  & $2.79^{+0.53}_{-0.81}$ & $2.99^{+0.35}_{-0.35}$ &
                  $9.14^{+0.96}_{-0.96}$ & $12$ \\
 A2052          & 0.0348  & $1.43^{+0.16}_{-0.12}$ & $1.62^{+0.03}_{-0.03}$ &
                  $3.21^{+0.15}_{-0.15}$ & $3.6$ \\
 EMSS  & 0.328   & $6.34^{+2.95}_{-2.86}$ & $7.37^{+5.36}_{-2.09}$ &
                  $8.46^{+1.93}_{-1.42}$ & $8.0$ \\
 \hline
 \end{tabular}
 \end{center}
 \caption{\footnotesize{In the list of clusters, Sersic stands for Sersic 159-03
 and EMSS for EMSS 1358+6425. 
 Redshift $z$, time-asymptotic central temperature
$T_f$, inner observed temperature $T_{inner}$, outer observed
temperature $T_{outer}$, and input initial temperature $T_i$, all in keV. 
The errors are given at the 1$ \sigma$ level.
 }}
 \label{tab.b}
 \end{table}
%
%
\begin{figure}[ht]
\begin{center}
\vspace{-10cm} \epsfig{file=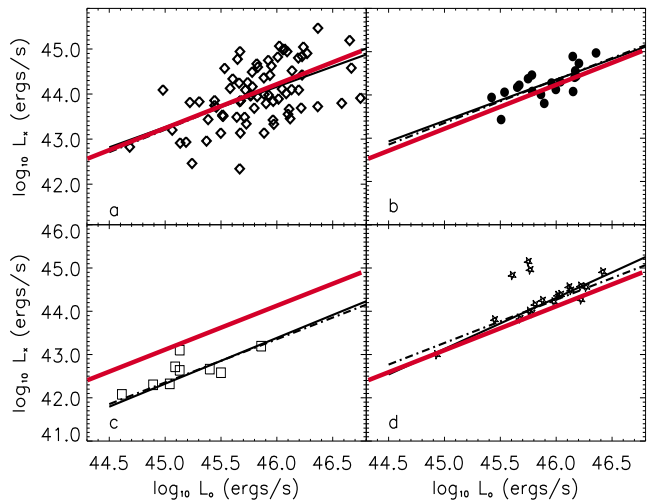,height=17.cm,width=11.cm,angle=0.0}
\end{center}
 \caption{\footnotesize{The X-ray versus the optical luminosity
of various sets of clusters, adopted from Miller and Nichol (2001). The thin (black)
lines are best fits of the form $L_X=a\, L_\star^\alpha$; in the dashed lines
$\alpha=1$ is enforced. The thick (red) line is the prediction of
Eq.~(\ref{lumcr}), for which $a\sim 0.02$ and $\alpha=1$. (a) is the Abell/ACO
sample, for which $\alpha=0.90\pm{0.17}$. (b) is the EDCC sample, for which
$\alpha=0.95\pm{0.22}$. (c) is the CfA {\it group} sample, for which
$\alpha=1.06\pm{0.11}$. (d) is the RASS sample, for which
$\alpha=1.18\pm{0.08}$. }} \label{lum}
\end{figure}
%
%
\begin{figure}[ht]
\vspace{7cm}
\epsfig{file=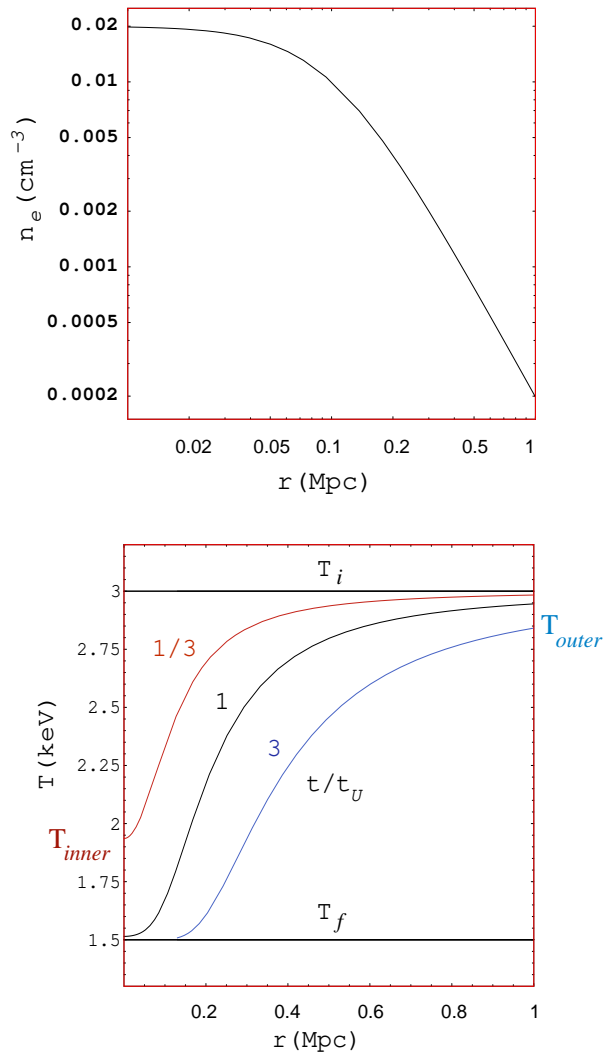,height=14.cm,width=8.cm,angle=0.0}
 \caption{\footnotesize{a) A typical CF-cluster density
profile. b) The time evolution of the temperature profile of the cluster, for
the given density profile, a typical WR luminosity and an initial
$T_i=3$ keV. The cluster ages correspond to 1/3, 1 and 3 times the
current age of the Universe. The meaning of the four temperatures,
$T_i$, $T_f$, $T_{inner}$ and $T_{outer}$, is illustrated.}} 
\label{fig.cfevolution}
\end{figure}
%
\begin{figure}[ht]
\begin{center}
\vbox{
 \epsfig{file=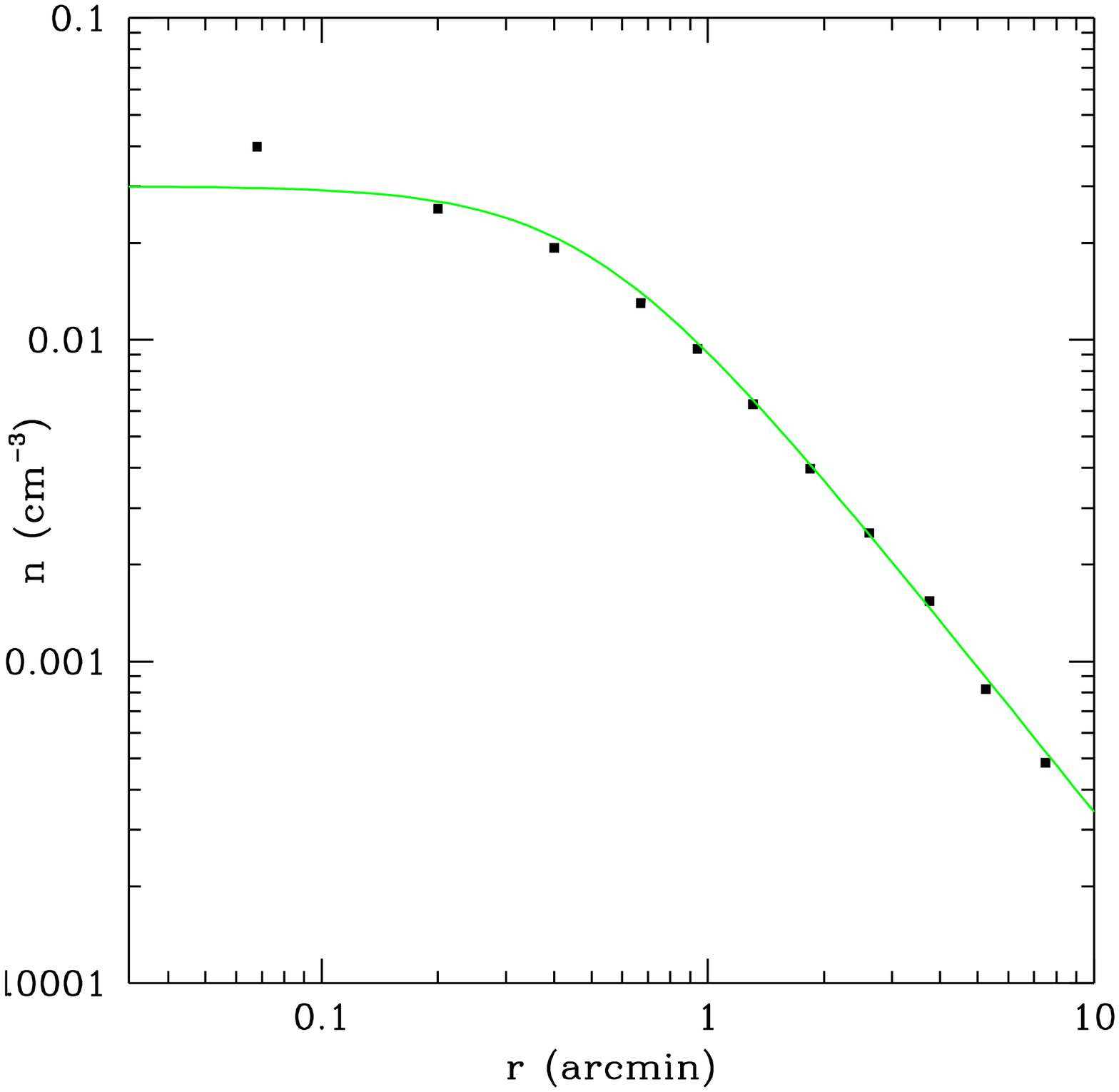,height=8.cm,width=8.cm,angle=0.0}
 \epsfig{file=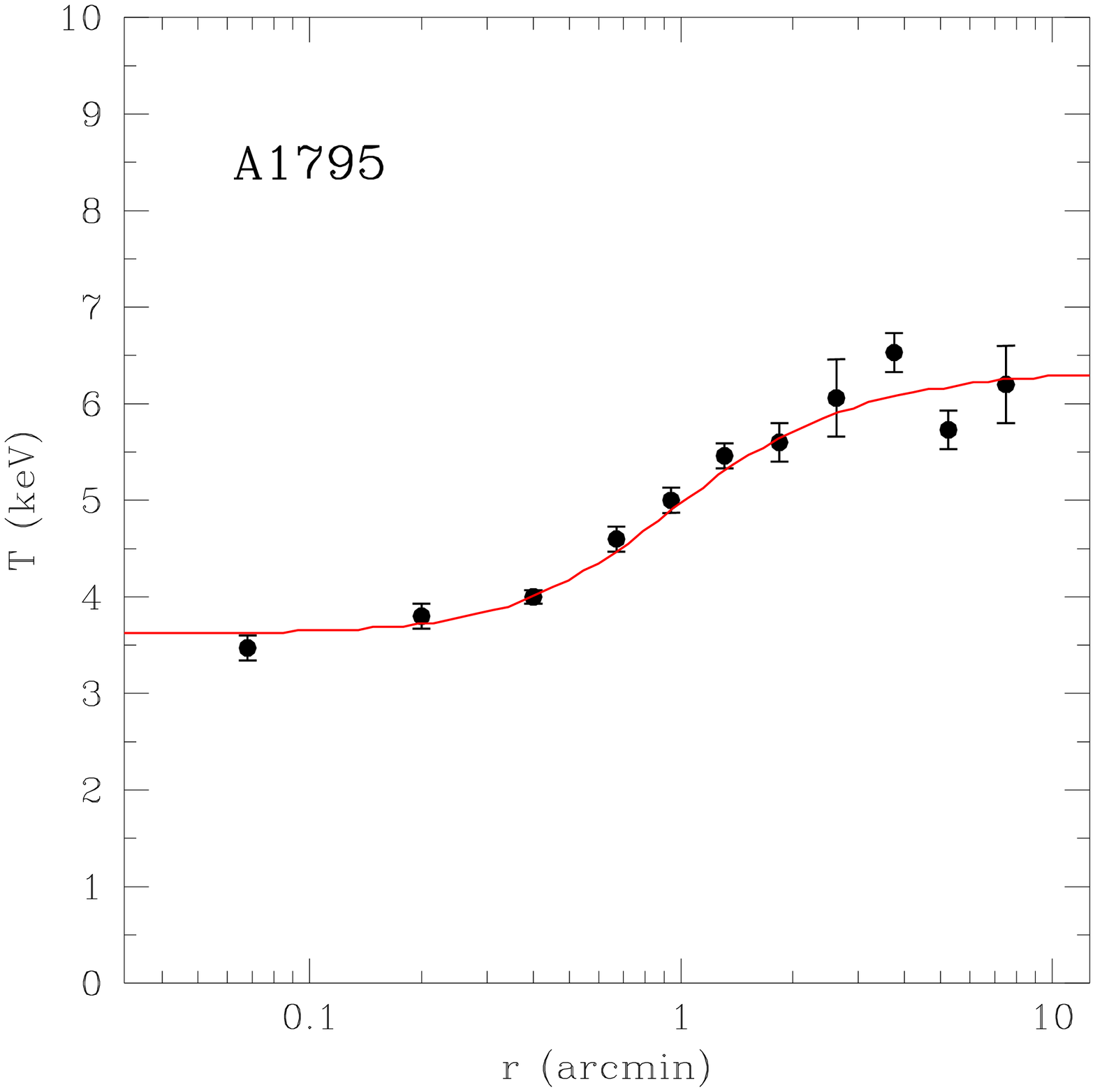,height=8.cm,width=8.cm,angle=0.0}
}
\end{center}
 \caption{\footnotesize{a)  The density profile 
of the cluster A1795. b) Its predicted temperature distribution. The data are
from Tamura et al. (2001).}} 
\label{fig.1795}
\end{figure}
\begin{figure}[ht]
\begin{center}
\vbox{
 \epsfig{file=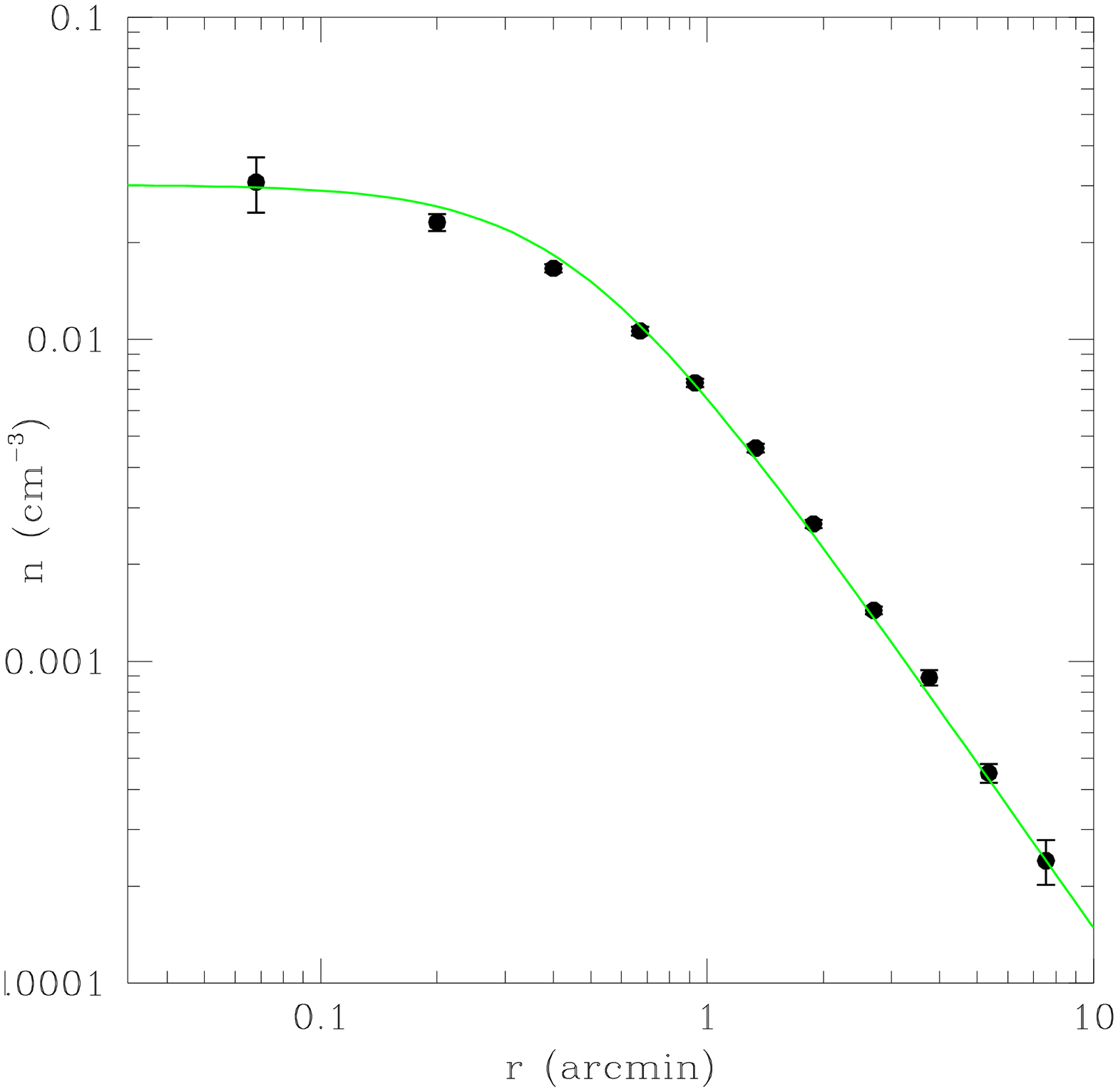,height=8.cm,width=8.cm,angle=0.0}
 \epsfig{file=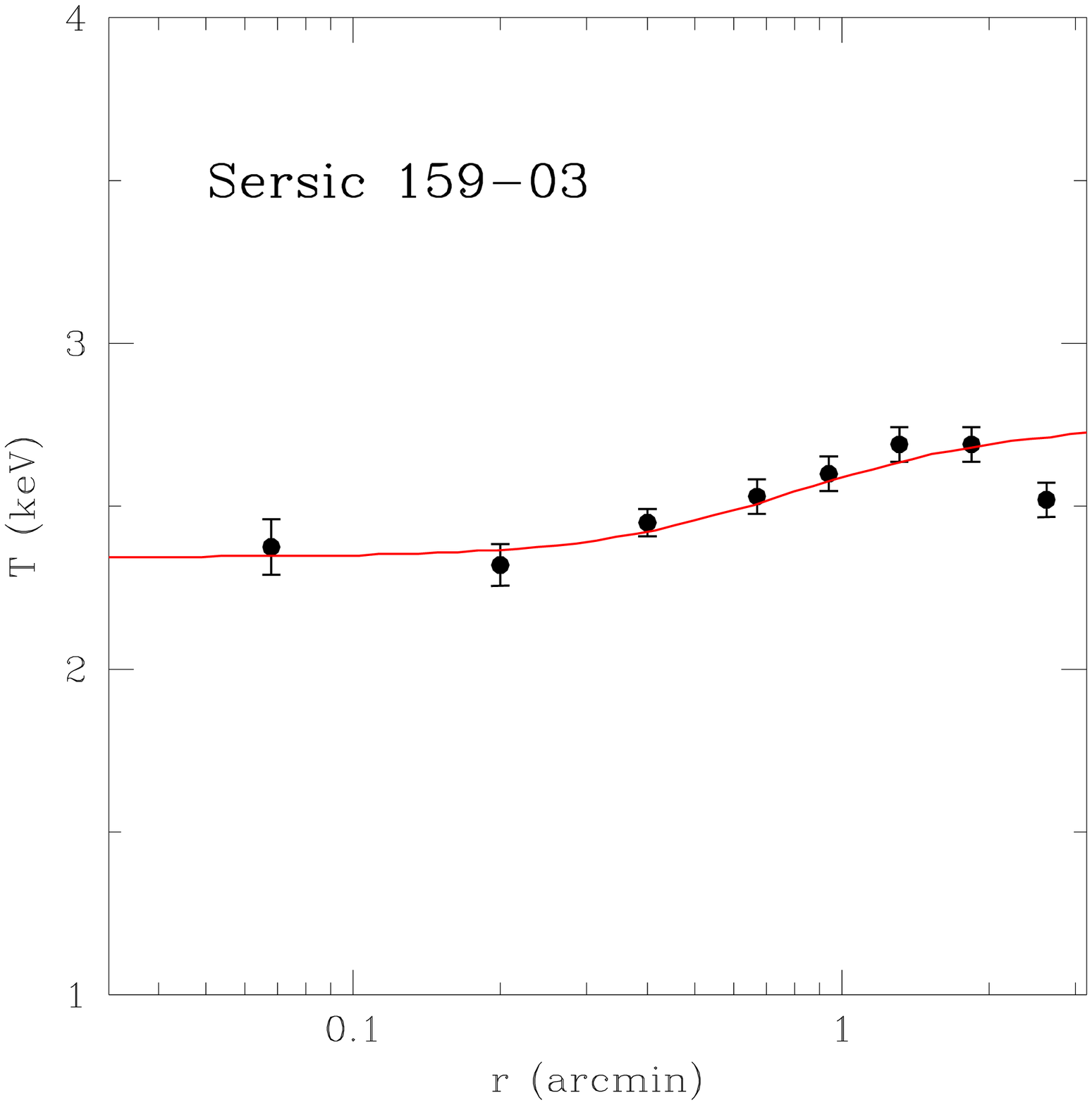,height=8.cm,width=8.cm,angle=0.0}
}
\end{center}
 \caption{\footnotesize{a) The density profile of the cluster
Sersic 159-03. b) Its predicted temperature distribution. The data are from
Kaastra et al. (2001). The $T_{outer}$ value in Table I is that of the penultimate
radial point.}} 
\label{fig.Sersic}
\end{figure}
\begin{figure}[ht]
\begin{center}
\vbox{
 \epsfig{file=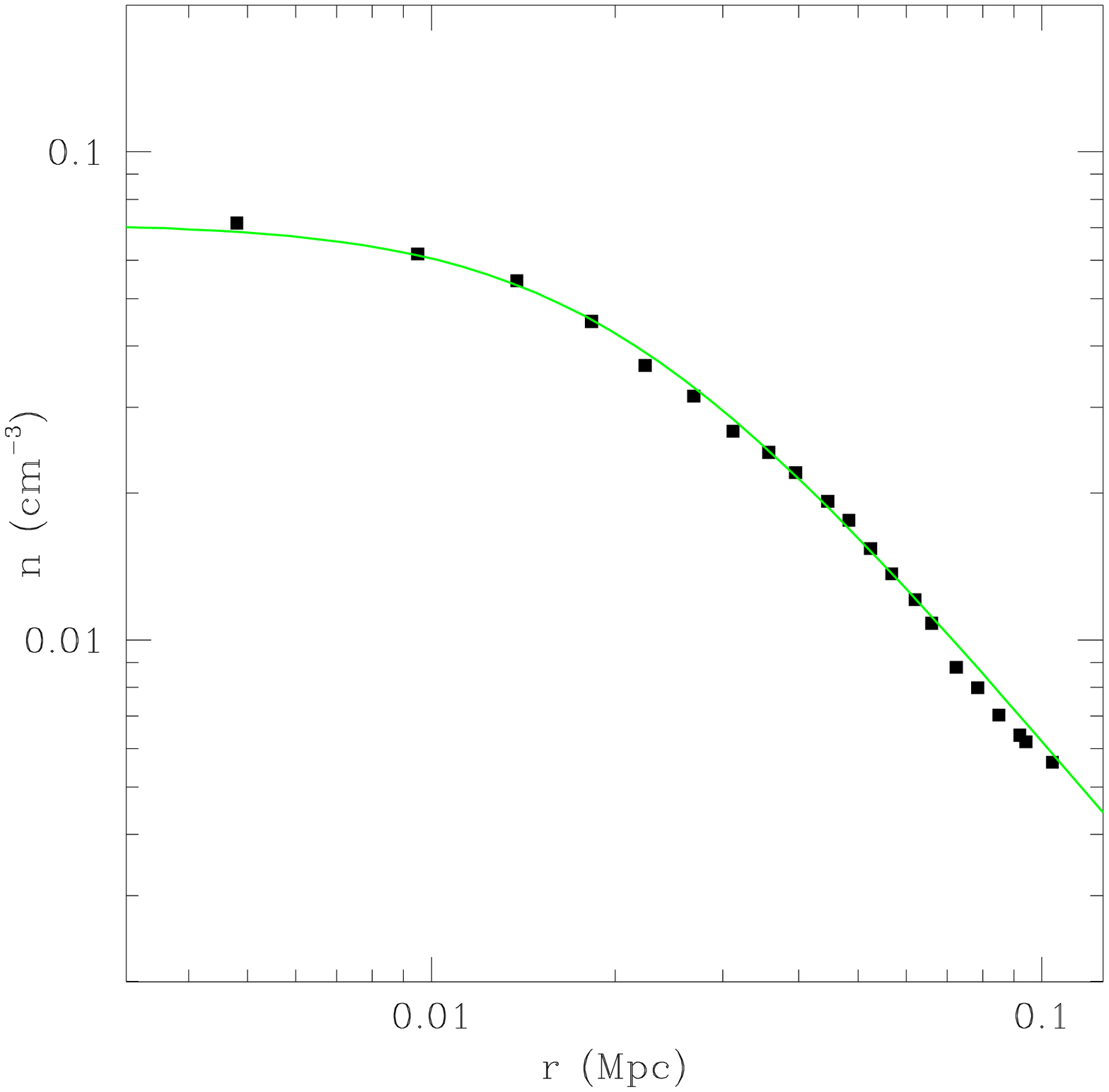,height=8.cm,width=8.cm,angle=0.0}
 \epsfig{file=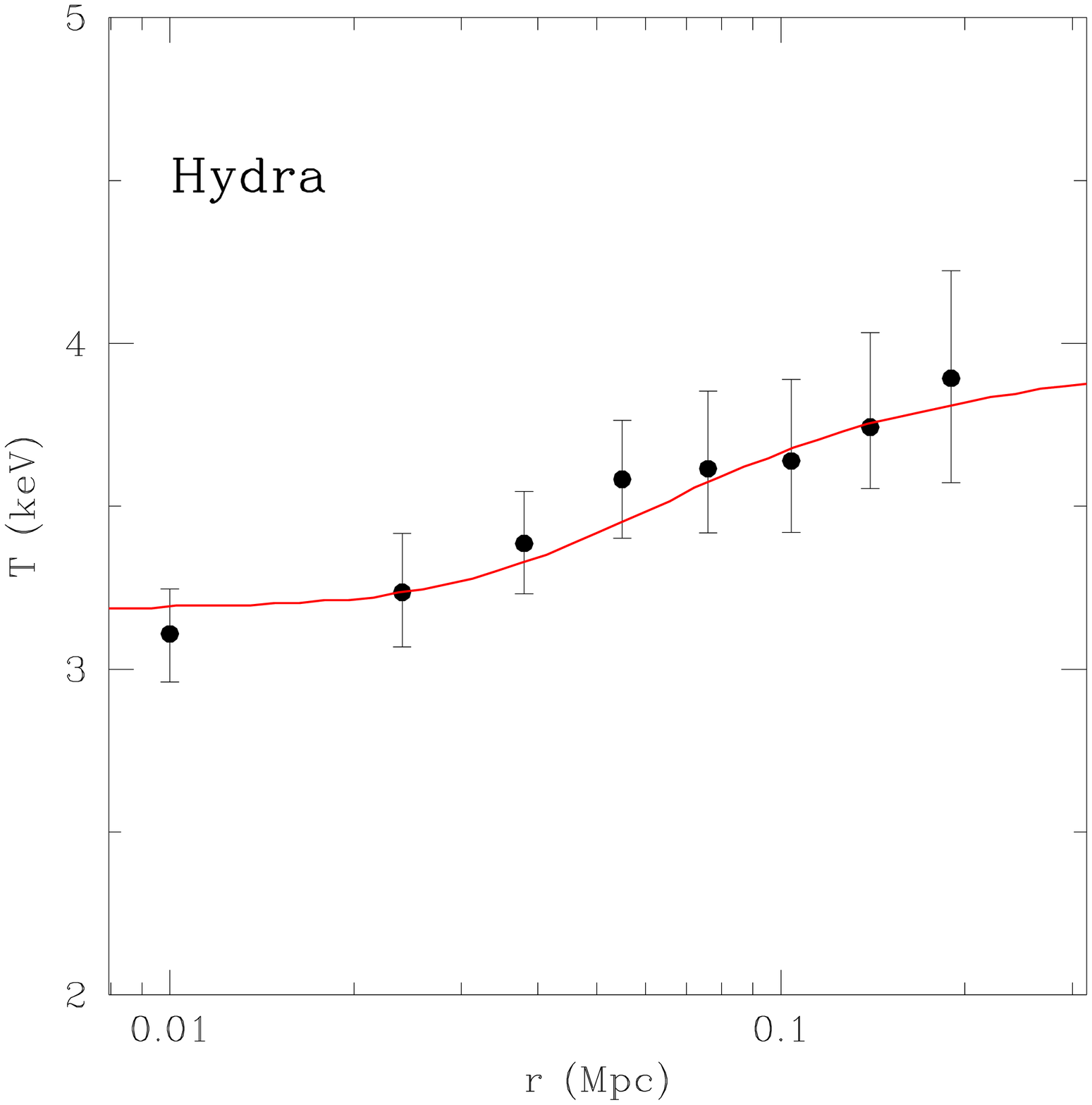,height=8.cm,width=8.cm,angle=0.0}
}
 \end{center}
 \caption{\footnotesize{a) The density profile of the cluster
Hydra. b) Its predicted temperature distribution. The data are from David et al.
(2001).}} 
\label{fig.Hydra}
\end{figure}
\begin{figure}[ht]
 \begin{center}
\vbox{
 \epsfig{file=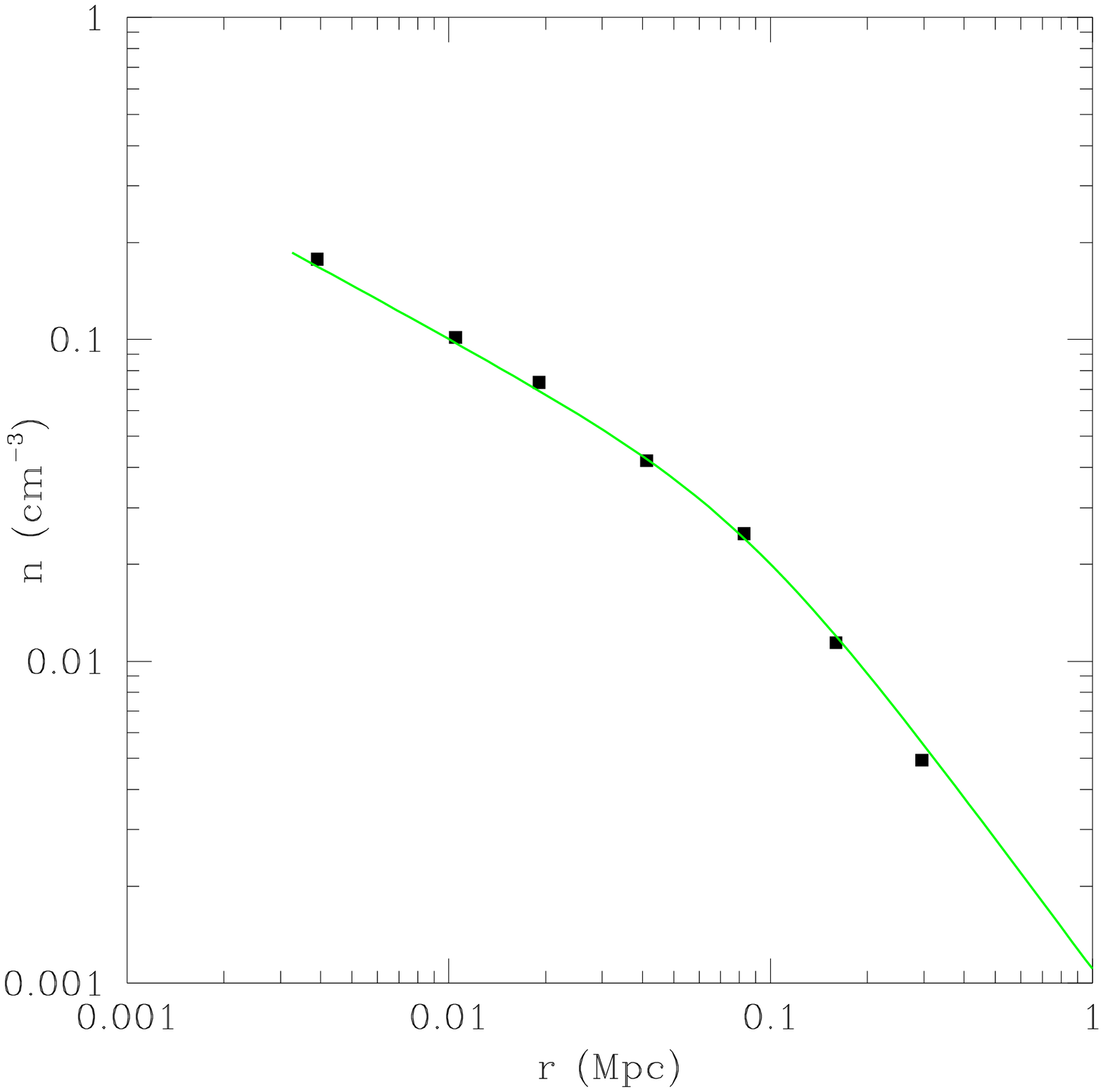,height=8.cm,width=8.cm,angle=0.0}
 \epsfig{file=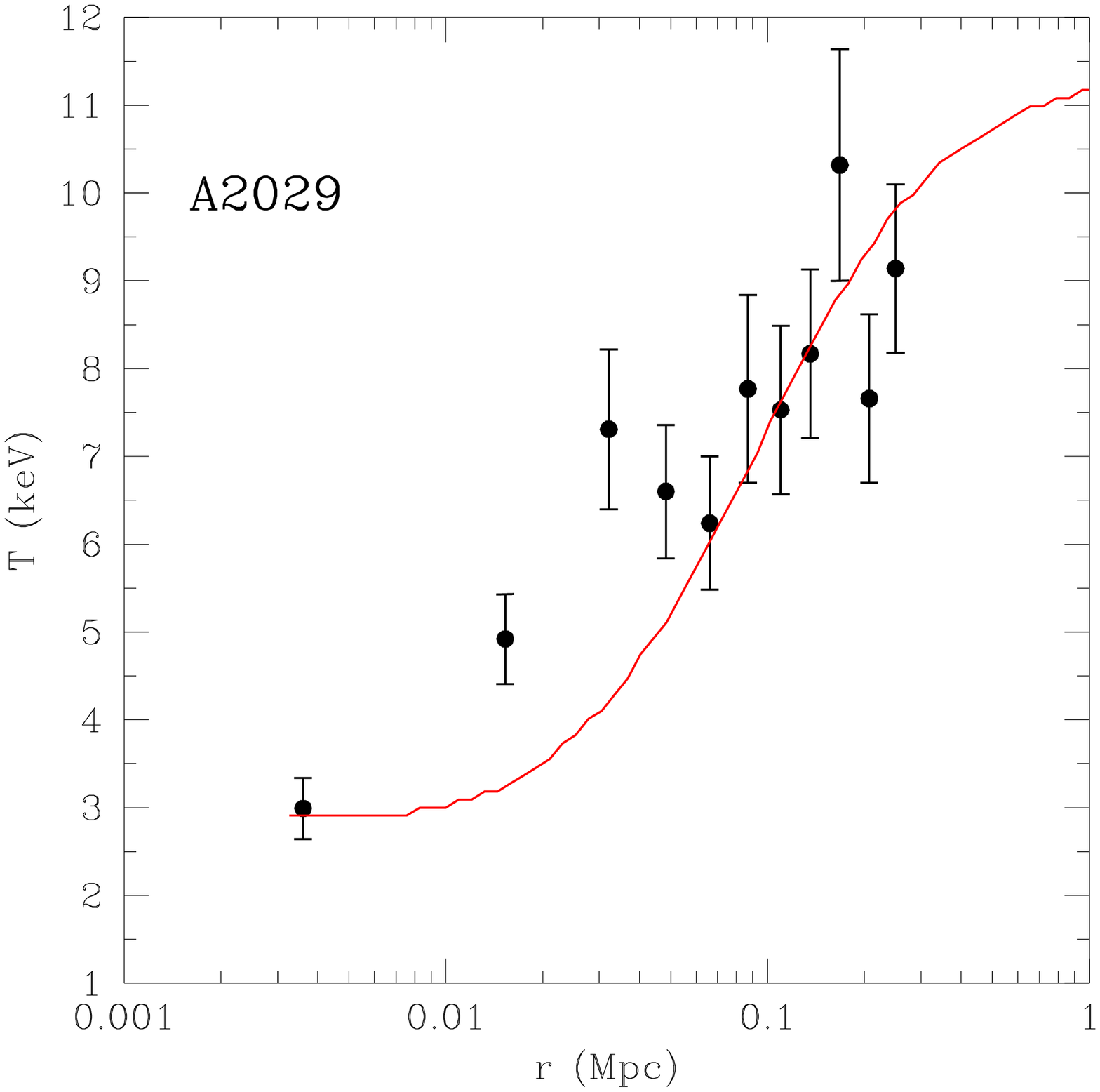,height=8.cm,width=8.cm,angle=0.0}
}
 \end{center}
 \caption{\footnotesize{a) The density profile of the
cluster A2029. b) Its predicted temperature distribution. The data are from
Lewis et al. (2002).}}
 \label{fig.2029}
\end{figure}
\begin{figure}[ht]
\begin{center}
\vbox{
 \epsfig{file=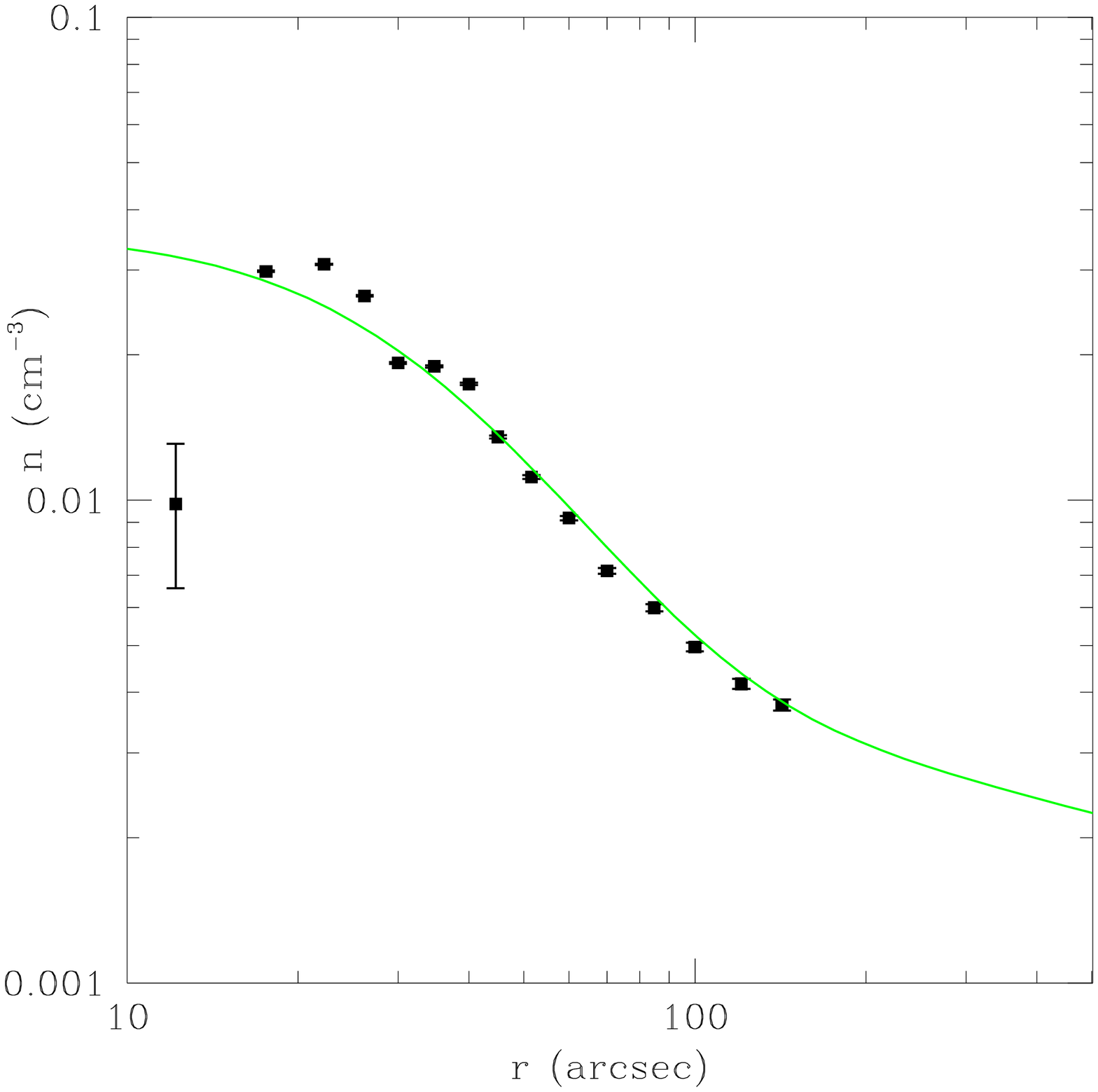,height=8.cm,width=8.cm,angle=0.0}
 \epsfig{file=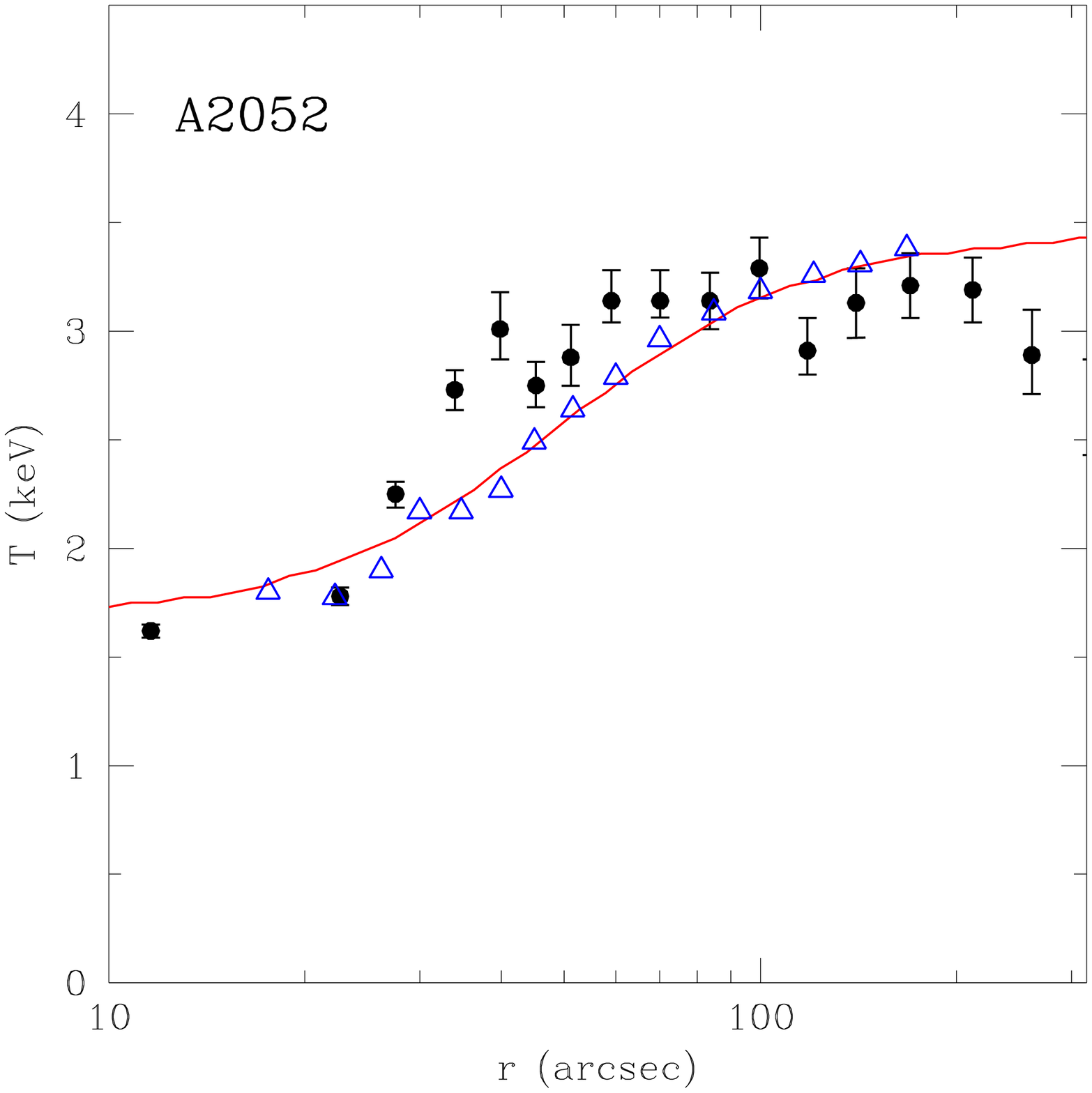,height=8.cm,width=8.cm,angle=0.0}
}
 \end{center}
 \caption{\footnotesize{a) The density profile of the
cluster A2052. b) Its predicted temperature distribution. The data are from
Blanton et al. (2002). The $T_{outer}$ value in Table I is that of the penultimate
radial point.}}
 \label{fig.2052}
\end{figure}
\begin{figure}[ht]
\begin{center}
\vbox{
 \epsfig{file=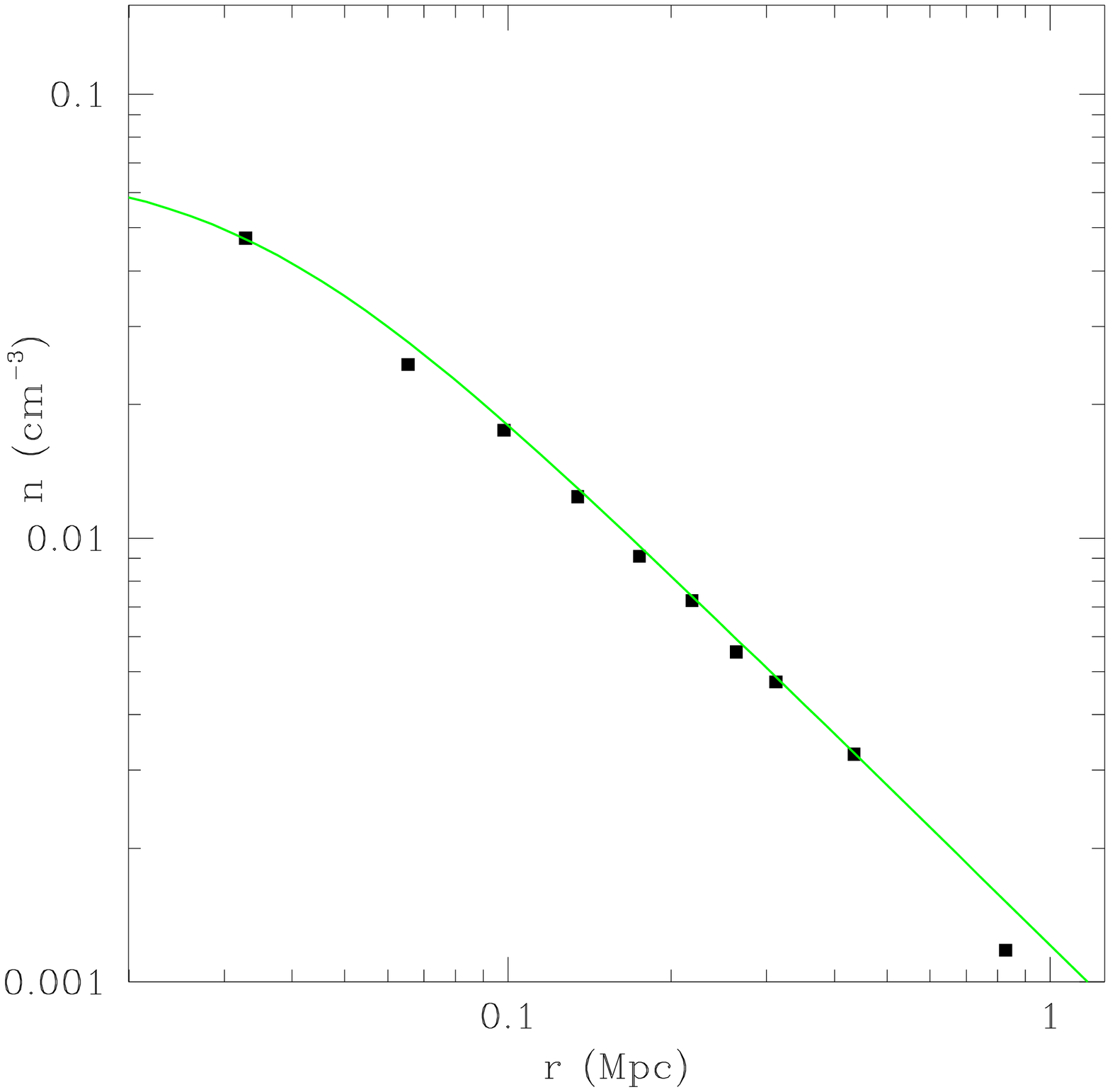,height=8.cm,width=8.cm,angle=0.0}
 \epsfig{file=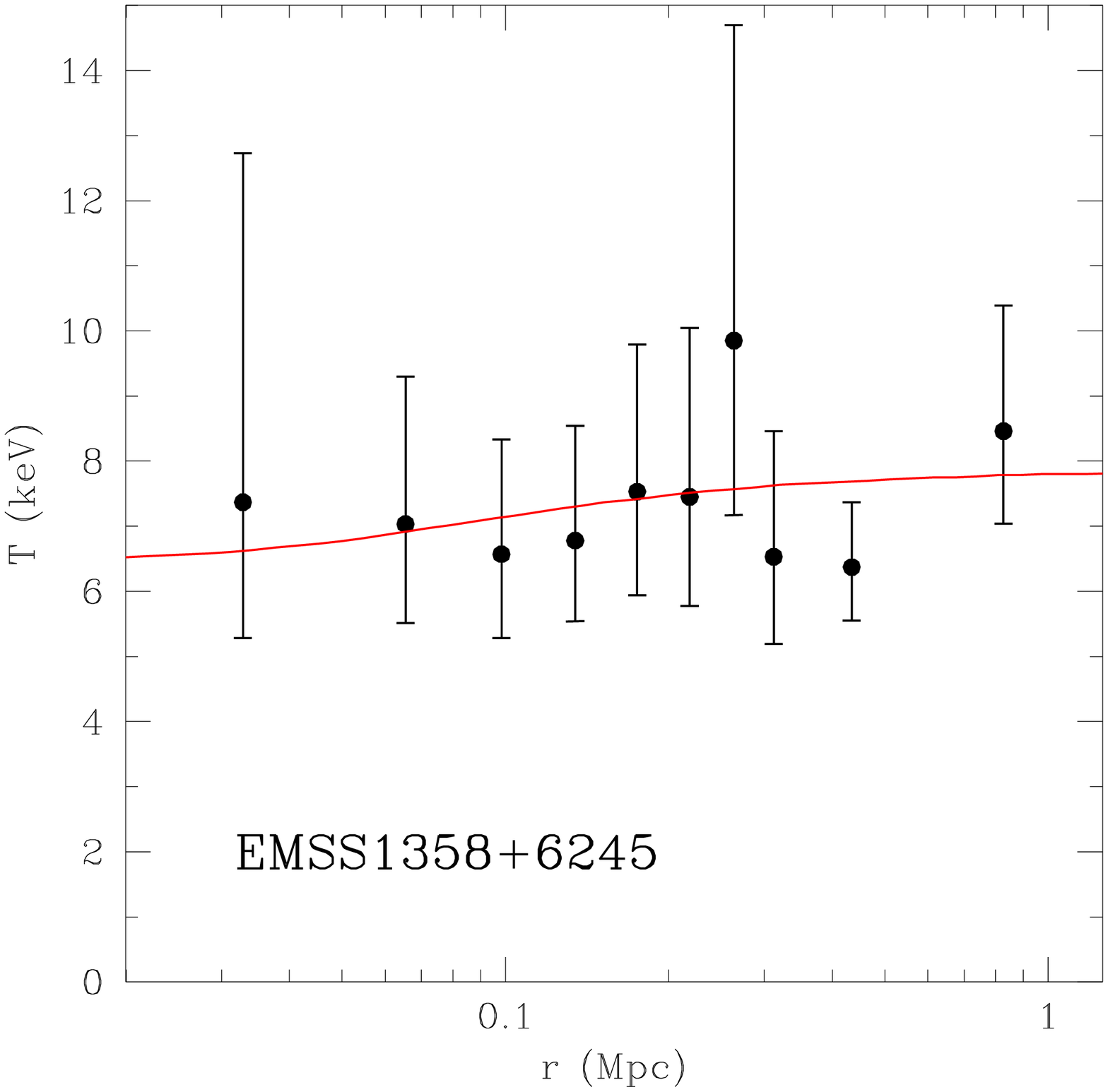,height=8.cm,width=8.cm,angle=0.0}
}
\end{center}
 \caption{\footnotesize{a) The density profile of the
cluster EMSS1358+6425. Its predicted temperature distribution. The data are from
Arabadjis et al. (2002).}}
 \label{fig.emss}
\end{figure}
\begin{figure}[ht]
\vspace{1cm} \epsfig{file=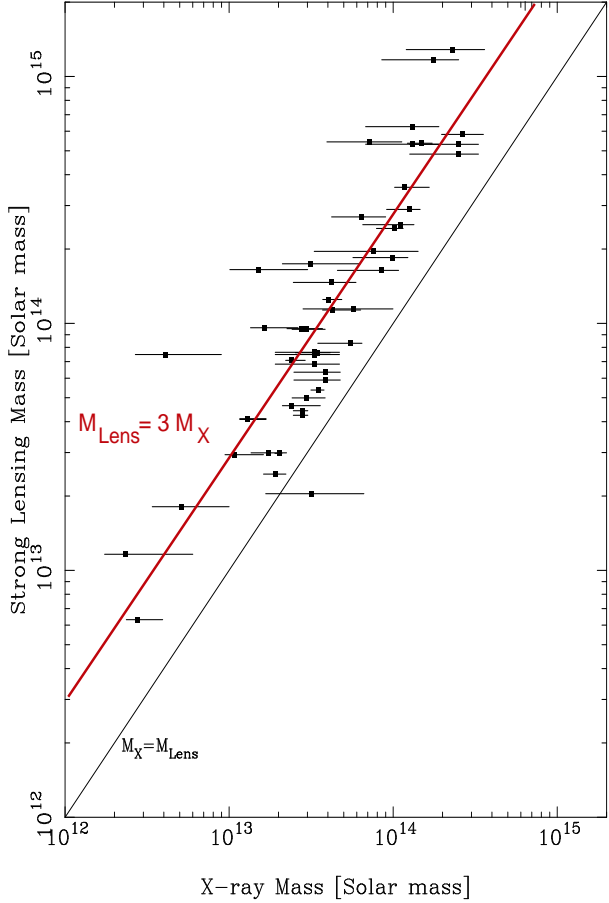,height=12.cm,width=8.cm,angle=0.0}
 \caption{\footnotesize{a) The lensing and X-ray masses of a selection
of clusters, from Hattori et al. (1999). The thinner (black) line is for equal
masses and the wider (red) line is $M_{Lens}=3\, M_X$, the result for
``equipartition'' of the plasma, the WRs and the magnetic field.}}
\label{mass}
\end{figure}
\begin{figure}[ht]
\begin{center}
\vspace{1cm}
\vbox{\epsfig{file=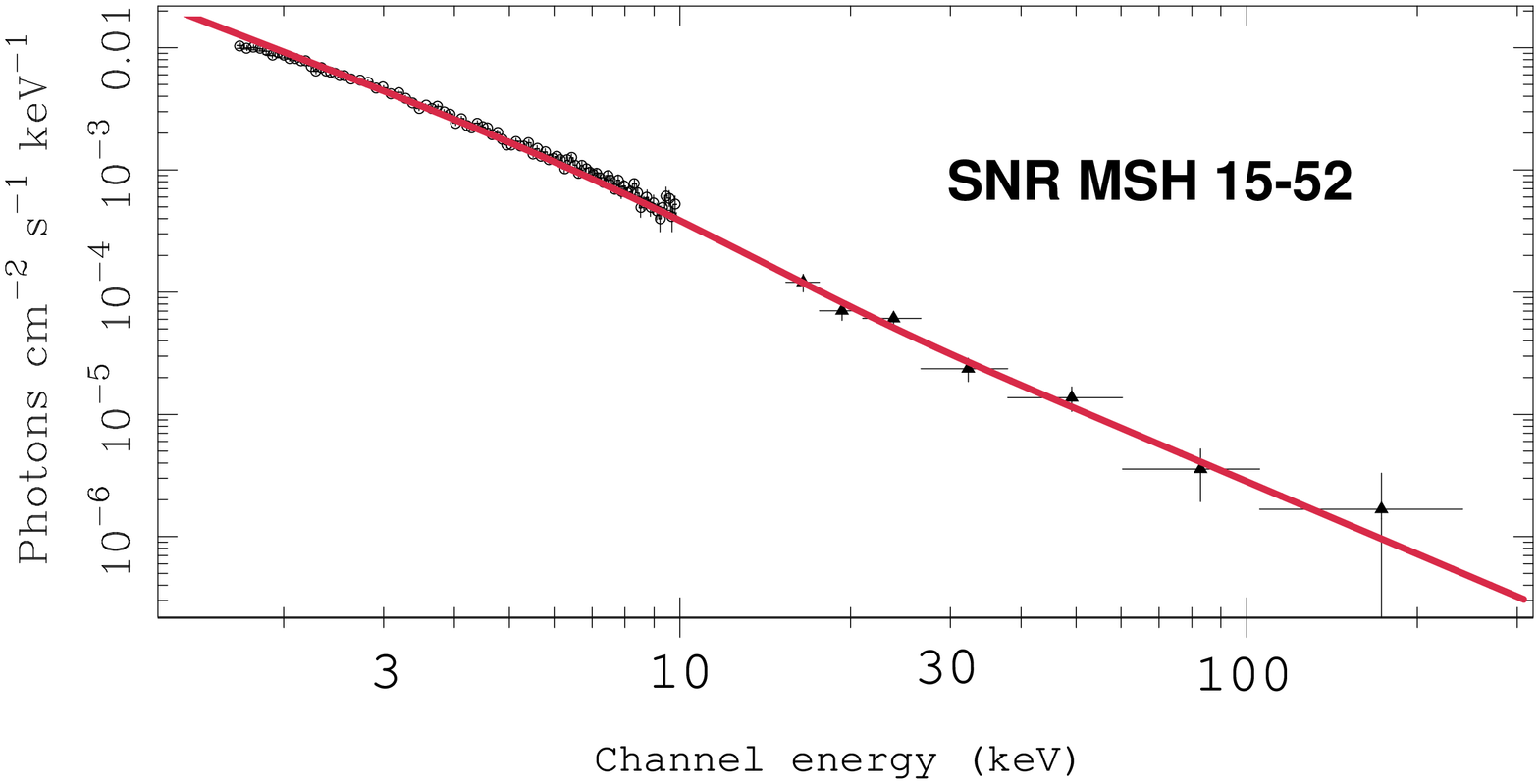,height=10.7cm,width=8.cm,angle=-90}}
\vspace{-2.5cm}
\vbox{\epsfig{file=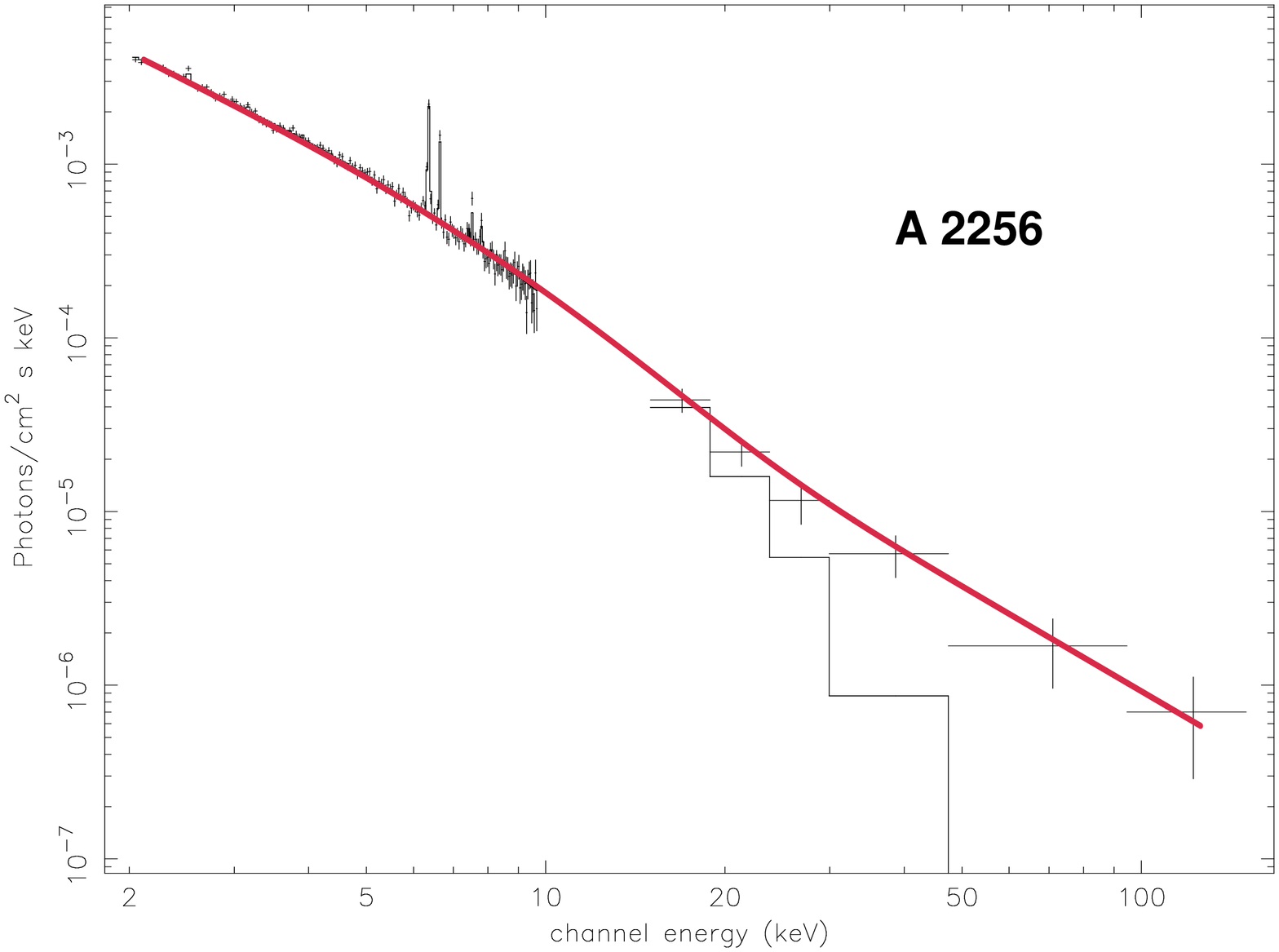,height=11cm,width=8.cm,angle=-90}}
\vspace{-1cm}
\end{center}
 \caption{\footnotesize{Upper panel: The thick (red) line is a
``$\chi$-by-eye'' fit of Eq.~(\ref{dndt}) to the thin thermal bremsstrahlung
emission from the SN remnant SNR MSH 15-52. The data and figure are from Mineo
et al. (2001). Lower panel: The same for the cluster A2256. The data and figure
are from Fusco-Femiano et al. (2000). The thin (black) line is their 
binned, purely thermal extrapolation.}}
 \label{spectrum}
\end{figure}

\end{document}